\documentclass[reprint]{revtex4-1}
\usepackage{amsfonts}
\usepackage{mathrsfs}
\usepackage{gensymb}
\usepackage{bbm}
\usepackage{dsfont}

\usepackage{chemformula}

\usepackage{soul}
\usepackage{xcolor}

\definecolor{webgreen}{rgb}{0,.5,0}
\definecolor{webbrown}{rgb}{.6,0,0}
\definecolor{grigio}{rgb}{.85,.85,.85} 
\definecolor{RoyalBlue}{rgb}{0.0, 0.14, 0.4}
\definecolor{skyblue1}{rgb}{0.45,0.62,0.81}
\definecolor{skyblue2}{rgb}{0.2,0.39,0.64}
\definecolor{skyblue3}{rgb}{0.13,0.29,0.53}
\definecolor{scarlet1}{rgb}{0.93,0.16,0.16}
\definecolor{scarlet2}{rgb}{0.8,0,0}
\definecolor{scarlet3}{rgb}{0.64,0,0}

\usepackage{hyperref}
\hypersetup{%
    colorlinks=true, linktocpage=true, pdfstartpage=1, pdfstartview=FitV,%
    breaklinks=true, pdfpagemode=UseNone, pageanchor=true, pdfpagemode=UseOutlines,%
    plainpages=false, bookmarksnumbered, bookmarksopen=true, bookmarksopenlevel=1,%
    hypertexnames=true, pdfhighlight=/O,
    urlcolor=webbrown, linkcolor=RoyalBlue, citecolor=webgreen, 
    pdftitle={Thermodynamics of Chemical Waves},%
    pdfauthor={Francesco Avanzini},%
    pdfsubject={},%
    pdfkeywords={},%
    pdfcreator={pdfLaTeX},%
    pdfproducer={LaTeX REVTeX}%
}

\DeclareMathAlphabet{\mathpzc}{OT1}{pzc}{m}{it}

\begin{document}

\title{Thermodynamics of Chemical Waves}
\newcommand\unilu{\affiliation{Complex Systems and Statistical Mechanics, Physics and Materials Science Research Unit, University of Luxembourg, L-1511 Luxembourg}}
\author{Francesco Avanzini}
\unilu
\author{Gianmaria Falasco}
\unilu
\author{Massimiliano Esposito}
\unilu

\date{\today}

\begin{abstract}

Chemical waves constitute a known class of dissipative structures emerging in reaction-diffusion systems. They play a crucial role in biology, spreading information rapidly to synchronize and coordinate biological events. We develop a rigorous thermodynamic theory of reaction-diffusion systems to characterize chemical waves. Our main result is the definition of the proper thermodynamic potential of the local dynamics as a nonequilibrium free energy density and establishing its balance equation. This enables us to identify the dynamics of the free energy, of the dissipation, and of the work spent to sustain the wave propagation. Two prototypical classes of chemical waves are examined. From a thermodynamic perspective, the first is sustained  by relaxation towards equilibrium and the second by nonconservative forces generated by chemostats. We analytically study step-like waves, called wavefronts, using the Fisher-Kolmogorov equation as representative of the first class and oscillating waves in the Brusselator model as representative of the second. Given the fundamental role of chemical waves as message carriers in biosystems, our thermodynamic theory constitutes an important step toward an understanding of information transfers and processing in biology.
\end{abstract}

\maketitle
\section{Introduction}
Chemical waves, or traveling waves, are complex concentration patterns of chemical species moving in space with a constant velocity and without changes of shape~\cite{Murray2002}.
They are essential for communication in biosystems both at the intracellular and intercellular level, and play a crucial role for the synchronization and coordination of biological events. 
By using chemical reactions, chemical waves are able to spread signals more rapidly compared to simple diffusion~\cite{Deneke2018}. 
The information they carry is encoded not only in the identity of the chemical species but also in other features of the concentration patterns such as the amplitude for step-like waves, called wavefronts, or the wavenumber for periodic waves~\cite{Purvis2013, Behar2010}. 
Calcium waves, for instance, are highly versatile intracellular messenger creating different spatiotemporal patterns~\cite{Berridge2003, Berridge2000} that regulates several cellular activities over a wide range of time scales~\cite{Carafoli2001}.
 

From a thermodynamic standpoint, chemical waves are out of equilibrium processes requiring a continual influx of energy and chemicals. They belong to the broader class of dissipative structures occurring in reaction-diffusion systems~\cite{Prigogine1971, Nicolis1977}.  
A lot of work has been dedicated to investigating their origin and their relative stability, but mainly from a dynamical point of view~\cite{Cross1993a, Ross2008}. However, a proper understanding of the energetic cost needed to support chemical waves is missing. 
We fill this gap in this work by developing a thermodynamic theory of chemical waves.


We start in Sec.~\ref{sec_rds} by providing a local formulation of thermodynamics of reaction-diffusion systems.
Our theory is based on recent advances in the thermodynamic description of chemical reaction networks~\cite{Rao2016, Rao2018b} and reaction-diffusion systems~\cite{Falasco2018a} inspired by stochastic thermodynamics~\cite{Jarzynski2011, VanDenBroeck2015} and its links to information theory~\cite{Parrondo2015}.
In Ref.~\cite{Falasco2018a}, global thermodynamic quantities were defined in terms of the reaction-diffusion dynamics to study the energetics of dissipative structures. However, to charaterize localized processes such as chemical waves, a local formulation is needed in the spirit of the phenomenological nonequilibrium thermodynamics developed in the last century~\cite{Groot1984, Glansdorff1971}. The theory we develop does so but without relying on any linear approximation of the thermodynamic currents in the forces. It is systematically built on  top of the dynamics and thus valid arbitrarily far from equilibrium. 
The thermodynamic quantities are local and expressed in terms of densities.   
A crucial step is the identification of the proper thermodynamic potential for the local dynamics, i.e., a free energy density.  
In a way which is reminiscent of equilibrium thermodynamics when passing from canonical to grand canonical ensembles, this potential is constructed from the Gibbs free energy density by eliminating the energetic contribution due to the local diffusion of the chemical species and by making use of the conservation laws of the chemical reaction network~\cite{Polettini2014, Rao2018a, Rao2018b}. Its balance equation provides a local formulation of the second law and decomposes the evolution of the free energy density into a flow of free energy caused by diffusion and three source terms. The first two source terms are chemical works produced respectively by time dependent driving and by nonconservative forces while the third one is a sink term due to entropy production.     
For undriven detailed balanced systems (i.e., in absence of work), this free energy  is minimized  as the system relaxes to equilibrium. 

  
We proceed in Sec.~\ref{sec_tw} by specializing our theory to study the thermodynamic cost of propagating and sustaining chemical waves. 
We first identify the conditions needed for an open reaction diffusion system to allow the existence of chemical waves. 
We then show in general that while the free energy is exclusively changing due to propagation, work and entropy production also quantify the cost for sustaining the waves.
We subsequently consider in Sec.~\ref{sec_ms} two prototypical wave patterns which we characterize analytically, i.e., wavefronts in the Fisher-Kolmogorov equation~\cite{Fisher1937, Murray2002} and periodic/oscillating waves in the Brusselator model~\cite{Prigogine1968, Auchmuty1976}. 
We find that when wavefronts are caused by relaxation towards equilibrium in detailed balanced systems, as Fisher-Kolmogorov waves, they propagate efficiently from an energetic standpoint because dissipation is solely localized at the front. Instead, oscillating waves, as in the Brusselator model, are significantly more energy costly as they need to be sustained by nonconservative forces (non detailed balanced systems) which cause dissipation throughout the system. 


The implications of our work for information processing in biosystems as well as possible future developments are discussed in Sec.~\ref{sec_conc}.



\section{Reaction-Diffusion Systems\label{sec_rds}}
A Reaction-Diffusion System (RDS) is considered here as a dilute ideal mixture of chemical species $\ch{Z}_{\alpha}$ ($\alpha\in\mathcal S$) diffusing within a volume $V$ with impermeable boundaries $\partial V$ and undergoing elementary reactions $\rho\in\mathcal R$. Each chemical species $\alpha$ is classified either as a controlled/chemostatted species ($\alpha_y\in Y$) if it is exchanged with particular reservoirs called chemostats controlling its concentration or as internal species $(\alpha_x\in X)$ otherwise~\cite{Rao2016}. It is characterized by its concentration field $Z^{\alpha}(\boldsymbol r,t)$ which follows the reaction-diffusion equation:
\begin{equation}
\partial_tZ^\alpha=\mathbb S^\alpha_\rho j^\rho - \nabla\cdot \boldsymbol J^\alpha +\delta^\alpha_{\alpha_y}I^{\alpha_y},
\label{eq_reaction_diffusion}
\end{equation}
where $\delta^\alpha_{\alpha_y}$ is the Kronecker delta and repeated upper-lower indices imply the summation over all the allowed values of the indices in all the paper. $j^\rho$ gives the concentration variation due to the chemical reaction $\rho$, $\mathbb S^\alpha_{+\rho}\ch{Z}_{\alpha}\mathbb \rightleftharpoons \mathbb S^\alpha_{-\rho}\ch{Z}_{\alpha}$. The stoichiometric coefficient $\mathbb S^\alpha_{+\rho}$ ($\mathbb S^\alpha_{-\rho}$) specifies the number of molecules $\alpha$ involved in the forward reaction $+\rho$ (backward reaction $-\rho$) whose chemical reaction current $j^{+\rho}$ ($j^{-\rho}$) satisfy the mass-action kinetics~\cite{Groot1984, Pekar2005}, \begin{equation}
j^{\pm\rho}=k_{\pm\rho}\prod_\alpha (Z^{\alpha})^{\mathbb S^{\alpha}_{\pm\rho}}.\label{eq_mass_action}
\end{equation}
The so-called stoichiometric matrix $\mathbb S^\alpha_\rho = \mathbb S^\alpha_{-\rho} - \mathbb S^\alpha_{+\rho}$ gives the full variation of the number of molecules $\alpha$ upon the reaction $\rho$, while $ j^\rho= j^{+\rho}-j^{-\rho}$ specifies the net reaction current. The Fick's diffusion currents
\begin{equation}
\boldsymbol J^{\alpha}= - D^{\alpha} \nabla Z^{\alpha},
\end{equation}
with the diffusion coefficient $D^{\alpha}$ for the species $\alpha$, describe the transport of chemical species within the volume $V$. They vanish at the boundaries $\partial V$, i.e., $\int_V\mathrm dr\text{ }\nabla\cdot\boldsymbol J^{\alpha}=0$ $\forall \alpha$. If the RDS is open, the controlled species $\alpha_y\in Y$ are exchanged with the chemostats according to the external currents $I^{\alpha_y}(\boldsymbol r,t)$. These latter vanish instead in a closed RDS.

The left null eigenvectors of the stoichiometric matrix,
\begin{equation}
l^\lambda_\alpha\mathbb S^\alpha_\rho=0\text{ }\text{ }\forall \rho\in\mathcal R,
\end{equation}
are the so-called conservation laws $l^\lambda$~\cite{Polettini2014, Rao2016} defining the quantities $L^\lambda =\int_V\mathrm d\boldsymbol r \text{ }l^\lambda_\alpha Z^\alpha$ that are conserved if the RDS is closed~\cite{Falasco2018a}: $\mathrm d_t L^\lambda=0$. When the system is open, the set of conservation laws $l^\lambda$ is split into two disjoint subsets: the unbroken conservation laws $l^{\lambda_u}$ and the broken conservation laws $l^{\lambda_b}$. The unbroken conservation laws are left null eingenvectors of the submatrix of the internal species, namely $l^{\lambda_u}_{\alpha_x}\mathbb S^{\alpha_x}_\rho=0$ and $l^{\lambda_u}_{\alpha_y}=0$, whereas the broken conservation laws are not, namely $l^{\lambda_b}_{\alpha_x}\mathbb S^{\alpha_x}_\rho\neq0$ for at least one $\rho\in\mathcal R$. Therefore, the quantities
\begin{equation}
L^{\lambda_u} =\int_V\mathrm d\boldsymbol r \text{ }l^{\lambda_u}_{\alpha_x} Z^{\alpha_x}
\label{eq_unbroken_conserved_quantities}
\end{equation}
are conserved even if the RDS is open, $\mathrm d_t L^{\lambda_u}=0$, unlike $L^{\lambda_b}=\int_V\mathrm d\boldsymbol r \text{ }l^{\lambda_b}_{\alpha} Z^{\alpha}$. They are called unbroken conserved quantities. Notice that chemostating a species does not always break a conservation law~\cite{Rao2016}. We thus distinguish the set of controlled species $\alpha_{y_b}\in Y_b\subseteq Y$ breaking all the broken conservation laws from the others $\alpha_{y_p}\in Y_p= Y \setminus Y_b$. This allows us to introduce the so-called moieties 
\begin{equation}
M^{\alpha_{y_b}}:=\hat{l}^{\alpha_{y_b}}_{\lambda_b}\int_V\mathrm d\boldsymbol r\text{ }l^{\lambda_b}_{\alpha}Z^{\alpha}
\label{eq_moieties}
\end{equation}
where $\hat{l}^{\alpha_{y_b}}_{\lambda_b}$ denotes the elements of the inverse of the matrix whose entries are $l^{\lambda_b}_{\alpha_{y_b}}$ (see Ref.~\cite{Falasco2018a} for details). The moieties represent the concentration of parts of molecules which are exchanged with the environment through the chemostats since their time evolution is determined only by the external currents, $\mathrm d_t M^{\alpha_{y_b}}(t)=\hat{l}^{\alpha_{y_b}}_{\lambda_b}\int_V\mathrm d\boldsymbol r\text{ }l^{\lambda_b}_{\alpha_y}I^{\alpha_y}(\boldsymbol r, t)$. 

The thermodynamic equilibrium is characterized by vanishing reaction currents $j^{\rho}=0$, diffusion currents $\boldsymbol J^{\alpha}=0$, and external currents $I^{\alpha_y}=0$. The equilibrium concentrations $Z_{\text{eq}}^{\alpha}$ are, consequently, homogeneously distributed over the volume $V$.  The existence of such equilibrium, together with mass-action kinetics, implies the so-called \textit{local detailed balance} condition for the kinetic constants $k_{\pm\rho}$ of the chemical reactions, 
\begin{equation}
k_{+\rho}/k_{-\rho}=\prod_\alpha (Z^{\alpha}_{\text{eq}})^{\mathbb S^{\alpha}_{\rho}}.
\end{equation}
This constraint is assumed to be valid to ensure that closed systems relax to equilibrium and hence thermodynamic consistency~\cite{Polettini2014}. Nonequilibrium conditions can be created by chemostatting certain species. These may be homogeneously distributed or not as well as constant in time or not. Our description of the reaction-diffusion dynamics presumes that all degrees of freedom other than concentrations in space to be equilibrated, as the temperature $T$ and the pressure of the solvent. In this way, thermodynamic state functions can be specified by their equilibrium form but expressed in terms of nonequilibrium concentrations, like for the chemical potential $\mu_{\alpha}(\boldsymbol r, t)={\mu}_{\alpha}^{\circ}+RT\ln( Z_{\alpha}(\boldsymbol r, t))$ of each species (${\mu}_{\alpha}^{\circ}$ is the standard chemical potential). In this respect, the local detailed balance condition can be restated establishing a correspondence between the kinetic constants $k_{\pm\rho}$, namely the dynamics of the RDSs, and the standard chemical potentials $\mu_\alpha^{\circ}$, namely the thermodynamics: 
\begin{equation}
\frac{k_{+\rho}}{k_{-\rho}}=\exp\left(-\frac{\mathbb S^{\alpha}_\rho\mu_\alpha^{\circ}}{RT}\right).
\label{eq_local_detailed_balance}
\end{equation}

We consider now the nonequilibrium thermodynamic characterization of RDSs. First, we recapitulate the global thermodynamic theory developed in Ref.~\cite{Falasco2018a} and then we generalize it to a local formulation. The global second law of thermodynamics for open RDSs can be stated as the following balance equation \begin{equation}
\mathrm d_t\mathcal G=-T\dot{\Sigma}+\dot{W}_{\text{driv}}^{\mathcal G}+\dot{W}_{\text{nc}},\label{eq_semigrand_balance_equation}
\end{equation}
for the so-called semigrand Gibbs free energy $\mathcal G(t)$. $T\dot{\Sigma}$ is the non-negative entropy production rate accounting for the energy dissipation, $\dot{W}_{\text{driv}}^{\mathcal G}$ is the driving work rate needed to manipulate the concentration of the chemostatted species in a time dependent way and $\dot{W}_{\text{nc}}$ is the nonconservative work rate spent to prevent the system from relaxing towards equilibrium. We examine in detail each term in Eq.~\eqref{eq_semigrand_balance_equation}. Whenever possible, we provide the expression of the thermodynamic quantities according to their densities which will then be used in the local formulation of thermodynamics. With a slight abuse of notation, we will use the same name for the density and the corresponding global quantities. For example, we will refer to both $T\dot{\sigma}$ and $T\dot{\Sigma}=\int_V\mathrm d\boldsymbol r \text{ }T\dot{\sigma}$ as entropy production rate. 

In analogy to equilibrium thermodynamics when passing from canonical to grand canonical ensembles, the semigrand Gibbs free energy
\begin{equation}
\mathcal G:=G-\mu^{\text{ref}}_{\alpha_{y_b}}M^{\alpha_{y_b}}\label{eq_semigrand_gibbs_free_energy}
\end{equation}
is obtained from the Gibbs free energy $G$ by eliminating the energetic contributions of the matter exchanged with the reservoirs.  The latter amounts to the moieties  $M^{\alpha_{y_b}}$ of Eq.~\eqref{eq_moieties}, times the reference values of their chemical potentials $\mu^{\text{ref}}_{\alpha_{y_b}}$ which are the values of chemical potential fixed by the chemostats $\alpha_{y_b}$. If a chemostat $\alpha_{y_b}$ sets different values of the chemical potential $\mu_{\alpha_{y_b}}(\boldsymbol r)$ for different points $\boldsymbol r$ in the volume $V$, then the reference chemical potential can be chosen arbitrarily among these values. This is equivalent to set the minimum value of $\mathcal G$ as it will become clear later. The Gibbs free energy $G(t)=\int_{V}\mathrm d\boldsymbol r\text{ }g(\boldsymbol r, t)$ is the integral over the volume $V$ of the Gibbs free energy density $g(\boldsymbol r, t)$ of ideal dilute solutions 
\begin{equation}
g=Z^{\alpha}\mu_\alpha -RT Z^{\mathcal S}\label{eq_gibbs_free_energy_density}
\end{equation}
with $RTZ^{\mathcal S}=RT\sum_{\alpha}Z^\alpha$ accounting for the contribution of the solvent~\cite{Ge2016, Fermi1956}.

The total entropy production rate $\dot{\Sigma}$ consists of two parts $\dot{\Sigma}=\dot{\Sigma}_{\text{rct}}+\dot{\Sigma}_{\text{diff}}$, i.e., the reaction $\dot{\Sigma}_{\text{rct}}$ and the diffusion $\dot{\Sigma}_{\text{diff}}$ part: 
\begin{align}
&\dot{\Sigma}_{\text{rct}}=\int_V\mathrm d\boldsymbol r\text{ }\dot{\sigma}_{\text{rct}}\geq0, &\dot{\Sigma}_{\text{diff}}=\int_V\mathrm d\boldsymbol r\text{ }\dot{\sigma}_{\text{diff}}\geq0.
\label{eq_global_entropy_production}
\end{align}
with
\begin{align}
&T\dot{\sigma}_{\text{rct}}=-(\mu_{\alpha}\mathbb S^{\alpha}_{\rho})j^{\rho}\geq0, &T\dot{\sigma}_{\text{diff}}= -(\nabla \mu_{\alpha})\cdot\boldsymbol J^{\alpha}\geq0.
\label{eq_entropy_production_rate}
\end{align}
The free energy of reaction $-(\mu_{\alpha}\mathbb S^{\alpha}_{\rho})$ and the variation of the chemical potential across space $-\nabla \mu_{\alpha}$ are thermodynamic forces driving the system towards equilibrium. The densities of the entropy production rates, as well as their global counterparts, are non-negative quantities and they vanish only at equilibrium. Indeed, the reaction entropy production rate can be written as $T\dot{\sigma}_{\text{rct}}=RTj^{\rho}\ln(j_{+\rho}/j_{-\rho})\ge 0$, with the aid of the local detailed balance~\eqref{eq_local_detailed_balance} and the mass-action kinetics~\eqref{eq_mass_action}. Similarly, the diffusion  entropy production rate can be written as $T\dot{\sigma}_{\text{diff}}=RTD^{\alpha}|\nabla Z_{\alpha}|^2/Z_{\alpha}\ge 0$. The driving work rate takes into account the time-dependent manipulation of $\mu_{\alpha_{y_b}}^{\text{ref}}$ performed by the chemostats $\alpha_{y_b}$
\begin{equation}
\dot{W}^{\mathcal G}_{\text{driv}}=-\mathrm d_t\mu^{\text{ref}}_{\alpha_{y_b}}M^{\alpha_{y_b}}.
\end{equation}
It obviously vanishes in autonomous systems. The nonconservative work rate
\begin{equation}
\dot{W}_{\text{nc}}=\int_{V}\mathrm d\boldsymbol r \text{ }\dot{w}_{\text{nc}},
\end{equation}
where the corresponding density is given by
\begin{equation}
\dot{w}_{\text{nc}}=\mathscr{F}_{\alpha_y}I^{\alpha_y},\label{eq_nonconservative_work_rate}
\end{equation}
quantifies the energetic cost of sustaining fluxes of chemical species among the chemostats by means of the forces $\mathscr{F}_{\alpha_y} = (\mu_{\alpha_y}-\mu^{\text{ref}}_{\alpha_{y_b}}\hat{l}^{\alpha_{y_b}}_{\lambda_b}l^{\lambda_b}_{\alpha_y})$. These forces have different origin depending on whether we consider the chemostatted species $\alpha_{y_b}$ or $\alpha_{y_p}$. Indeed, $\mathscr{F}_{\alpha_{y_b}}(\boldsymbol r)=\mu_{\alpha_{y_b}}(\boldsymbol r)-\mu^{\text{ref}}_{\alpha_{y_b}}$ is the difference of chemical potential of the same species imposed in different points of the space. It vanishes in case of homogeneous chemostatting.  In contrast, $\mathscr{F}_{\alpha_{y_p}}(\boldsymbol r)=\mu_{\alpha_{y_p}}(\boldsymbol r)-\mu^{\text{ref}}_{\alpha_{y_b}}\hat{l}^{\alpha_{y_b}}_{\lambda_b}l^{\lambda_b}_{\alpha_y}$ is the difference of chemical potentials imposed by different chemostats. It vanishes if all the chemostats break a conservation law, i.e., $Y_b=Y$.

In the absence of nonconservative forces $\mathscr{F}_{\alpha_y}=0$ $\forall \alpha_y\in Y$, no nonconservative work is performed $\dot{W}_{\mathrm nc}=0$ and the system is said to be \textit{detailed balanced}. This occurs when all the chemostats break a conservation law $Y_{b}=Y$, and the chemostatted species are homogeneously distributed $\mu_{\alpha_{y_b}}(\boldsymbol r)=\mu_{\alpha_{y_b}}^{\text{ref}}$ $\forall \boldsymbol r\in V$. If a detailed balanced system is also autonomous $\dot{W}_{\text{driv}}=0$, it relaxes towards equilibrium. This follows from the fact that the semigrand Gibbs free energy approaches its equilibrium value $\mathcal G_{\text{eq}}$ monotonously in time: i) the time derivative of $\mathcal G(t)$ is always negative, 
\begin{equation}
\mathrm d_t \mathcal G=-T\dot{\Sigma}\le 0.
\end{equation}
and ii) $\mathcal G(t)$ is lower bounded by its equilibrium value $ \mathcal G_{\text{eq}}$. Indeed, $\mathcal G$ can be expressed in terms of its reference equilibrium value $\mathcal G_{\text{eq}}$ as
\begin{equation}
\mathcal G = \mathcal G_{\text{eq}} + RT \int_V\mathrm d\boldsymbol r\text{ }\mathcal L(\{Z^{\alpha}(\boldsymbol r, t)\}\parallel\{Z^{\alpha}_{\text{eq}}\})
\end{equation}
by employing the relative entropy for the concentration fields (i.e., non-normalized concentration distributions) 
\begin{equation}
\mathcal L(\{Z^{\alpha}\}\parallel\{Z^{\alpha}_{\text{eq}}\})=\sum_{\alpha\in\mathcal S}Z^{\alpha}\ln\left(\frac{Z^{\alpha}}{Z^{\alpha}_{\text{eq}}}\right)-\left(Z^{\alpha} - Z^{\alpha}_{\text{eq}}\right),\label{eq_relative_entropy}
\end{equation}
which quantifies the difference between the two concentration distributions $\{Z^{\alpha}\}$ and $\{Z^{\alpha}_{\text{eq}}\}$. Since the relative entropy is always positive, unless $Z^{\alpha} = Z^{\alpha}_{\text{eq}}$, the semigrand Gibbs free energy is greater than or equal to its equilibrium counterpart $\mathcal G\geq \mathcal G_{\text{eq}}$. In other words, $\mathcal G(t)$ acts as a Lyapunov function. The reference equilibrium condition 
\begin{equation}
Z^{\text{eq}}_\alpha=\exp((\mu^{\text{ref}}_\alpha-\mu_\alpha^{\circ})/RT)\label{eq_equilibrium_condition}
\end{equation}
is specified according to the reference chemical potentials $\mu^{\text{ref}}_\alpha$. We have already introduced the reference chemical potentials  $\mu^{\text{ref}}_{\alpha_{y_b}}$ for the controlled species $\alpha_{y_b}$ in Eq.~\eqref{eq_semigrand_gibbs_free_energy}. Here, we call the chemical potentials $\mu_{\alpha_{x}}$ and $\mu_{\alpha_{y_b}}$ evaluated at the equilibrium condition reference chemical potentials because they depend on $\mu^{\text{ref}}_{\alpha_{y_b}}$. A detailed discussion in Appendix~\ref{app_ref_chem_pot} shows that, for the $Y_p$ controlled species, $\mu_{\alpha_{y_p}}^{\text{ref}}$ is given by $\mu_{\alpha_{y_p}}^{\text{ref}}=\mu_{\alpha_{y_b}}^{\text{ref}}\hat{l}^{\alpha_{y_b}}_{\lambda_b}l^{\lambda_b}_{\alpha_y}$. This also provides an interpretation of the thermodynamic force $\mathscr{F}_{\alpha_{y_p}}(\boldsymbol r)=\mu_{\alpha_{y_p}}(\boldsymbol r)-\mu^{\text{ref}}_{\alpha_{y_b}}\hat{l}^{\alpha_{y_b}}_{\lambda_b}l^{\lambda_b}_{\alpha_y}$ as the difference between the chemical potential of the species $\alpha_{y_b}$ imposed by the chemostat $\alpha_{y_b}$ and the corresponding reference chemical potential. In contrast, $\mu_{\alpha_{x}}^{\text{ref}}$ is specified by both the reference chemical potentials $\mu_{\alpha_{y_b}}^{\text{ref}}$ and the unbroken conserved quantities $L^{\lambda_u}$ of Eq.~\eqref{eq_unbroken_conserved_quantities} but an explicit expression cannot be obtained. If all the conservation laws are broken, there are no  more conserved quantities and also $\mu_{\alpha_{x}}^{\text{ref}}$ is given by $\mu_{\alpha_{x}}^{\text{ref}}=\mu_{\alpha_{y_b}}^{\text{ref}}\hat{l}^{\alpha_{y_b}}_{\lambda_b}l^{\lambda_b}_{\alpha_x}$. 
 It is import to understand that the choice of $\mu^{\text{ref}}_{\alpha_{y_b}}$ of Eq.~\eqref{eq_semigrand_gibbs_free_energy} set the equilibrium condition $\{Z^{\text{eq}}_\alpha\}$ reached by the system if it was detailed balanced. Consequently, the choice of $\mu^{\text{ref}}_{\alpha_{y_b}}$ set $\mathcal G_{\text{eq}}$ that is the minimum value of $\mathcal G(t)$.

We now generalize this description to a local thermodynamic theory. To this aim, a fundamental step is the identification of the proper thermodynamic potential characterizing the local dynamics of RDSs as $\mathcal G$ does for the global dynamics. Therefore, we seek a free energy density that i) is lower bounded by its equilibrium value and ii) approaches asymptotically its equilibrium value for undriven detailed balanced systems. One might wonder whether the semigrand Gibbs free energy density
\begin{equation}
\mathpzc g(\boldsymbol r,t):=g(\boldsymbol r,t)- \mu^{\text{ref}}_{\alpha_{y_b}}\hat{l}^{\alpha_{y_b}}_{\lambda_b}l^{\lambda_b}_{\alpha}Z^{\alpha}(\boldsymbol r, t)\label{eq_def_semigrand_gibss_free_energy_density},
\end{equation}
such that $\mathcal G(t)=\int_V\mathrm d\boldsymbol r\text{ }\mathpzc g(\boldsymbol r,t)$, plays this role. Its balance equation is specified as
 \begin{equation}
\partial_t\mathpzc g = -T\dot{\sigma}+\dot{w}_{\text{driv}}^{\mathpzc g}+\dot{w}_{\text{nc}}-\nabla\cdot\boldsymbol J^{\mathpzc g},
\label{eq_time_der_semigrand_gibss_free_energy_density}
\end{equation}
where $T\dot{\sigma}$ is the entropy production rate of Eq.~\eqref{eq_entropy_production_rate},  $\dot{w}_{\text{nc}}$ is the nonconservative work rate of Eq.~\eqref{eq_nonconservative_work_rate}, and  the driving work rate is given by 
\begin{equation}
\dot{w}_{\text{driv}}^{\mathpzc g} = -\mathrm d_t\mu^{\text{ref}}_{\alpha_{y_b}}\hat{l}^{\alpha_{y_b}}_{\lambda_b}l^{\lambda_b}_{\alpha}Z^{\alpha}.
\end{equation}
Compared to the evolution equation~\eqref{eq_semigrand_balance_equation} for the semigrand Gibbs free energy, a flow term 
\begin{equation}
\boldsymbol J^{\mathpzc g}=(\mu_{\alpha}-\mu^{\text{ref}}_{\alpha_{y_b}}\hat{l}^{\alpha_{y_b}}_{\lambda_b}l^{\lambda_b}_{\alpha})\boldsymbol J^{\alpha}
\end{equation}
arises at the local level. This term is absent at the global level because the diffusion currents vanish at the boundaries $\partial V$ and, consequently, $\int_V\mathrm d\boldsymbol r\text{ }\nabla\cdot\boldsymbol J^{\mathpzc g}=0$ .

The semigrand Gibbs free energy density $\mathpzc g(\boldsymbol r,t)$ is not the proper thermodynamic potential because it is not lower bounded by its equilibrium value $\mathpzc g_{\text{eq}}$: $\mathpzc g(\boldsymbol r,t)$ can not be written as $\mathpzc g_{\text{eq}}$ plus the relative entropy of Eq.~\eqref{eq_relative_entropy}. Consequently, $\mathpzc g(\boldsymbol r,t)$ is not minimized in undriven detailed balanced systems. This is a direct consequence of the definition~\eqref{eq_def_semigrand_gibss_free_energy_density} where $\mathpzc g(\boldsymbol r,t)$ is constructed from the Gibbs free energy density $g(\boldsymbol r, t)$ by eliminating the energetic contributions of the matter exchanged with the chemostats. Indeed, the term $\hat{l}^{\alpha_{y_b}}_{\lambda_b}l^{\lambda_b}_{\alpha}Z^{\alpha}(\boldsymbol r, t)$ in Eq.~\eqref{eq_def_semigrand_gibss_free_energy_density} is the local representation of the moieties of Eq.~\eqref{eq_moieties}. However, at the local level diffusion allows all the species to be exchanged with neighboring regions of space which play the same role as the chemostats from a local standpoint. This means that the proper thermodynamic potential must be specified as the following free energy density
\begin{equation}
\mathbbm g(\boldsymbol r,t) := g(\boldsymbol r, t)-\mu^{\text{ref}}_{\alpha}Z^\alpha(\boldsymbol r, t),
\label{eq_local_free_energy}
\end{equation}
where the energetic contributions of all the species exchanged through diffusion are removed from the Gibbs free energy density $g(\boldsymbol r, t)$ of Eq.~\eqref{eq_gibbs_free_energy_density}. The local equivalent to the term $\mu^{\text{ref}}_{\alpha_{y_b}}M^{\alpha_{y_b}}$ in Eq.~\eqref{eq_semigrand_gibbs_free_energy} is precisely $\mu^{\text{ref}}_{\alpha}Z^{\alpha}$. The reference chemical potentials $\mu^{\text{ref}}_\alpha$ have the same meaning as in Eq.~\eqref{eq_equilibrium_condition}. If all the conservation laws are broken by the chemostats, the definition of the free energy density $\mathbbm g(\boldsymbol r,t)$ in Eq.~\eqref{eq_local_free_energy} is equivalent to the definition of the semigrand Gibbs free energy density $\mathpzc g(\boldsymbol r,t)$ in Eq.~\eqref{eq_def_semigrand_gibss_free_energy_density} since $\mu^{\text{ref}}_{\alpha_x}Z^{\alpha}=\mu^{\text{ref}}_{\alpha_{y_b}}\hat{l}^{\alpha_{y_b}}_{\lambda_b}l^{\lambda_b}_{\alpha_x}Z^{\alpha_x}$ and $\mu^{\text{ref}}_{\alpha_{y_p}}Z^{\alpha}=\mu^{\text{ref}}_{\alpha_{y_b}}\hat{l}^{\alpha_{y_b}}_{\lambda_b}l^{\lambda_b}_{\alpha_{y_p}}Z^{\alpha_{y_p}}$. 

We can now verify that $\mathbbm g(\boldsymbol r, t)$ can be expressed in terms of the reference equilibrium free energy $\mathbbm g_{\text{eq}}$ as
\begin{equation}
\mathbbm g(\boldsymbol r, t) = \mathbbm g_{\text{eq}} + RT \mathcal L(\{Z^{\alpha}(\boldsymbol r, t)\}\parallel\{Z^{\alpha}_{\text{eq}}\})
\end{equation}
by employing the relative entropy of Eq.~\eqref{eq_relative_entropy}. The free energy density is therefore  always greater than or equal to the equilibrium counterpart $\mathbbm g\geq \mathbbm g_{\text{eq}}$. The fact that the relative entropy of information theory~\cite{cover2012} appears both at the global and the local level supports the idea that thermodynamics plays a fundamental role in the characterization of the information codified in complex patterns of chemical concentrations. 

Furthermore, the free energy density $\mathbbm g(\boldsymbol r,t)$ satisfies the balance equation 
\begin{equation}
\partial_t\mathbbm g = -T\dot{\sigma}+\dot{w}^{\mathbbm g}_{\text{driv}}+\dot{w}_{\text{nc}}-\nabla\cdot\boldsymbol J^{\mathbbm g}
\label{eq_time_der_local_free_energy}
\end{equation}
which constitutes the local formulation of the second law. The non-negative entropy production $T\dot{\sigma}$ and the nonconservative work rate $\dot{w}_{\text{nc}}$ are the same as in Eq~\eqref{eq_entropy_production_rate} and Eq.~\eqref{eq_nonconservative_work_rate}, respectively. However, the driving chemical work rate
\begin{equation}
\dot{w}_{\text{driv}}^{\mathbbm g} = -\mathrm d_t\mu^{\text{ref}}_{\alpha}Z^{\alpha}
\end{equation}
now takes into account the time evolution of all $\mu^{\text{ref}}_{\alpha}$ through the time-dependent manipulation of $\mu^{\text{ref}}_{\alpha_{y_b}}$ performed by the chemostats $\alpha_{y_b}$. Indeed, the reference chemical potentials $\mu^{\text{ref}}_{\alpha_{y_p}}$ and $\mu^{\text{ref}}_{\alpha_{x}}$ depend on $\mu^{\text{ref}}_{\alpha_{y_b}}$'s (see Appendix~\ref{app_ref_chem_pot}). The flow of free energy,
\begin{equation}
\boldsymbol{J}^{\mathbbm g} = (\mu_{\alpha}-\mu_{\alpha}^{\text{ref}})\boldsymbol J^{\alpha}
\label{eq_free_energy_diffusion}
\end{equation}
describes the effects on the evolution of the free energy density due to the local diffusion currents. Note that the diffusion entropy production rate $T\dot{\sigma}_{\text{diff}}$ (resp. the free energy flow $\boldsymbol J^{\mathbbm g}$) can be further split into a contribution due to the internal species $T\dot{\sigma}_{\text{diff}}^X=-(\nabla \mu_{\alpha_x})\cdot\boldsymbol J^{\alpha_x}$ (resp. $\boldsymbol{J}^{\mathbbm g_X} = (\mu_{\alpha_x}-\mu_{\alpha_x}^{\text{ref}})\boldsymbol J^{\alpha_x}$) and one due to the controlled species $T\dot{\sigma}_{\text{diff}}^Y=-(\nabla \mu_{\alpha_y})\cdot\boldsymbol J^{\alpha_y}$ (resp. $\boldsymbol{J}^{\mathbbm g_Y} = (\mu_{\alpha_y}-\mu_{\alpha_y}^{\text{ref}})\boldsymbol J^{\alpha_y}$). We will make use of this decomposition when we examine the thermodynamic description of chemical waves. Equations~\eqref{eq_local_free_energy} and~\eqref{eq_time_der_local_free_energy} constitute the first major result of this paper.

For undriven detailed balanced systems  $\mathbbm g$ approaches $\mathbbm g_{\text{eq}}$ asymptotically.
Indeed, the global free energy, 
\begin{equation}
\mathbbm G(t):=\int_V\mathrm d\boldsymbol r\text{ }\mathbbm g(\boldsymbol r,t),
\end{equation}
 is greater than its reference equilibrium value $\mathbbm G\geq\mathbbm G_{\text{eq}}$ (because $\mathbbm g\geq\mathbbm g_{\text{eq}}$) and its time derivative is always negative
\begin{equation}
\mathrm d_t\mathbbm G =-T\dot{\Sigma}\leq0
\end{equation}
since $\dot{w}^{\mathbbm g}_{\text{driv}}=\dot{w}_{\text{nc}}=0$ and $\int_{V}\mathrm d\boldsymbol r\text{ }\nabla\cdot\boldsymbol J^{\mathbbm g}=0$.  In other words, $\mathbbm G\to \mathbbm G_{\text{eq}}$ monotonously in time. While this implies that $\mathbbm g$ approaches $\mathbbm g_{\text{eq}}$ asymptotically, i.e., $\mathbbm g$ is minimized, it is not granted that its time derivative $\partial_t\mathbbm g= -T\dot{\sigma}-\nabla\cdot\boldsymbol J^{\mathbbm g}$ is always negative because of the free energy flow. Consider for example the case where the concentration for all the species except one is equal to its equilibrium value at one specific point $\boldsymbol r'$ in space: $Z^{\alpha}(\boldsymbol r, t)=Z^{\alpha}_{\text{eq}}$ $\forall \alpha\neq\alpha'$ if and only if $\boldsymbol r= \boldsymbol r'$. The concentration of the species $\alpha'$ differs instead from its equilibrium value for an arbitrarily small positive number, $Z^{\alpha'}(\boldsymbol r', t)-Z^{\alpha'}_{\text{eq}}=\epsilon>0$. Expanding the time derivative of the free energy $\partial_t\mathbbm g(\boldsymbol r, t)$ for $\boldsymbol r= \boldsymbol r'$ in powers of  $\epsilon$ and truncating the expansion at the lowest order gives $\partial_t\mathbbm g(\boldsymbol r', t)=\epsilon\mathrm d_\epsilon(T\dot{\sigma}_{\text{rct}}) + \epsilon D^{\alpha'}\nabla^2Z^{\alpha'}/Z^{\alpha'}_{\text{eq}}$, with the derivative of the reaction entropy production rate evaluated for $\epsilon=0$. Therefore, the free energy increases locally ($\partial_t\mathbbm g(\boldsymbol r', t)>0$) if, for instance, the curvature of the concentration field is very large at $\boldsymbol r = \boldsymbol r'$ (e.g., $\nabla^2Z^{\alpha'}=Z^{\alpha'}_{\text{eq}}/\epsilon^2$) since the contribution of the reaction entropy production rate becomes negligible.


\section{Chemical waves\label{sec_tw}}
We now examine the thermodynamic description of chemical waves considered here as particular pattern solutions of RDSs also called traveling waves. The concentration of a species $Z^{\alpha}(\boldsymbol r, t)$ evolves as a traveling wave if it propagates in space with a constant velocity $\boldsymbol c^{\alpha}$ and without changing its shape:
\begin{equation}
Z^{\alpha}(\boldsymbol r, t) = Z^{\alpha}(\boldsymbol r -\boldsymbol c^{\alpha} t).
\label{eq_traveling_wave_def}
\end{equation}
We label the spatial coordinate in the traveling wave $\alpha$ frame of reference as $\tilde{\boldsymbol r}^{\alpha}:={\boldsymbol r}-{\boldsymbol c}^{\alpha}t$.

\subsection{Dynamics of Traveling Waves\label{subsec_tw_dyn}}
We first discuss a set of conditions that the reaction-diffusion equation~\eqref{eq_reaction_diffusion} needs to satisfy to allow for traveling wave solutions. We will then employ these conditions in Subs.~\ref{subsec_tw_the} to specialize the general thermodynamic description of Sec.~\ref{sec_rds} for traveling waves. We start by splitting the chemical species into two disjoint subsets, i.e., the traveling waves $\alpha_x\in X$ and the other species $\alpha_y\in Y$. The traveling waves are internal species by construction. We use the same notation for the nontraveling wave species as for the controlled species in Sec.~\ref{sec_rds} since we will show  that these former must be undriven chemostatted species. 

Using Eq.~\eqref{eq_traveling_wave_def}, the reaction-diffusion equation~\eqref{eq_reaction_diffusion} for the traveling wave $\alpha_x$ becomes
\begin{equation}
D^{\alpha_x}\nabla^2 Z^{\alpha_x}+\boldsymbol c^{\alpha_x}\cdot\nabla Z^{\alpha_x} + \mathbb S^{\alpha_x}_{\rho}j^{\rho} =0.
\label{eq_rdequation_travwave}
\end{equation}
In the frame of reference of the traveling wave $\alpha_x$, the terms $\nabla^2 Z^{\alpha_x}$ and $\boldsymbol c^{\alpha_x}\cdot\nabla Z^{\alpha_x}$ are time-independent and, consequently, the above equation admits a solution if the concentration variation due to the chemical reactions is time-independent as well: $\partial_t(\mathbb S^{\alpha_x}_{\rho}j^{\rho}(\tilde{\boldsymbol r}^{\alpha_x} + \boldsymbol c^{\alpha_x}t, t))=0$. This condition must be satisfied for every $\tilde{\boldsymbol r}^{\alpha_x}$ and every $t$ justifying the hypothesis that each reaction current $j^\rho$ is time-independent in the frame of reference of the traveling wave $\alpha_x$, i.e., $\partial_t(j^{\rho}(\tilde{\boldsymbol r}^{\alpha_x} + \boldsymbol c^{\alpha_x}t, t))=0$ $\forall \rho\in\mathcal R$. In other words, each reaction current evolves like the traveling wave $\alpha_x$ with the propagation velocity $\boldsymbol c^{\alpha_x}$. Since this must be true for every traveling species, the reaction currents $j^\rho$ have to be time-independent in every frame of reference $\tilde{\boldsymbol r}^{\alpha_x}$.  Therefore, we conclude that all the traveling waves evolve with the same velocity $\boldsymbol c$ and the corresponding comoving coordinate is $\tilde{\boldsymbol r}=\boldsymbol r - \boldsymbol c t$.

This implies that the condition $\partial_t(j^{\rho}(\tilde{\boldsymbol r}+ \boldsymbol ct, t))=0$, using the mass-action kinetics Eq.~\eqref{eq_mass_action} and $\partial_t(Z^{\alpha_x}(\tilde{\boldsymbol r}))=0$, becomes
\begin{equation}
\sum_{\alpha_y\in Y}\frac{\partial j^{\rho}}{\partial Z^{\alpha_y}}\left[\boldsymbol c \cdot \nabla Z^{\alpha_y}+\partial_tZ^{\alpha_y}\right]=0,
\end{equation}
where $\nabla Z^{\alpha_y}$ and $\partial_t Z^{\alpha_y}$ are evaluated in frame of reference $\tilde{\boldsymbol r}$. The above equation must hold for every reaction current $j^\rho$ of every point $\tilde{\boldsymbol r}$ and of every time $t$. We thus assume that each term between square bracket vanish independently, $\boldsymbol c \cdot \nabla Z^{\alpha_y}+\partial_tZ^{\alpha_y}=0$. This means that the species $Y$ are either traveling waves $\partial_tZ^{\alpha_y}=-\boldsymbol c \cdot \nabla Z^{\alpha_y}$ or undriven controlled species $\partial_tZ^{\alpha_y}=0$ with a concentration field that is invariant under space translation along the propagation direction $\boldsymbol c \cdot \nabla Z^{\alpha_y}=0$. Since the species $Y$ are not traveling waves by hypothesis, they are undriven chemostatted species. In this respect, their reaction-diffusion equation simplifies to
\begin{equation}
\mathbb S^{\alpha_y}_{\rho}j^\rho-\nabla\cdot\boldsymbol J^{\alpha_y} + I^{\alpha_y}=0,
\label{eq_external_currents_tw}
\end{equation}
with vanishing diffusion currents in the propagation directions $\boldsymbol c \cdot \boldsymbol J^{\alpha_y}=-D^{\alpha_y}\boldsymbol c\cdot\nabla Z^{\alpha_y}=0$. 

In conclusion, to allow for traveling waves to possibly arise, we consider only RDSs that satisfy the following requirements:
\begin{itemize}
\item[(i)] the chemical species are split into two disjoint subsets, i.e., the traveling waves $X$ (internal species) and the controlled species $Y$;
\item[(ii)] the traveling waves are characterized by the same propagation velocity $\boldsymbol c$ and they have the same comoving coordinate $\tilde{\boldsymbol r} = \boldsymbol r -\boldsymbol c t$;
\item[(iii)] the concentration field of the controlled species $Y$ is invariant under space translation along the propagation direction $\boldsymbol c \cdot \nabla Z^{\alpha_y}=0$ and it is kept constant by the chemostats $\partial_tZ^{\alpha_y}=0$ (no driving work is performed, $\dot{w}_{\text{driv}}=0$).
\end{itemize}
Although we do not expect it, we note that rigorously speaking weaker conditions for traveling wave solutions may exist as our argumentation is not a mathematical proof.

\subsection{Thermodynamics of Traveling Waves\label{subsec_tw_the}}
We now examine how the thermodynamic quantities such as the local free energy, the entropy production rate, and the nonconservative work rate evolve when traveling waves propagate in RDSs.

Using Eq.~\eqref{eq_traveling_wave_def}, the time evolution of the free energy is given by
\begin{equation}
\partial_t\mathbbm g = -(\mu_{\alpha_x}-\mu^{\text{ref}}_{\alpha_x})\boldsymbol c\cdot\nabla Z^{\alpha_x},
\label{eq_time_der_local_free_energy_trav_wave}
\end{equation}
where one can recognize a drift contribution for every traveling wave $\boldsymbol c\cdot\nabla Z^{\alpha_x}$. In other words, the free energy is ``dragged'' by the propagating traveling waves in the $\boldsymbol c$ direction.

The nonconservative work rate and the entropy production rate account for the energetic cost of supporting the wave propagation. Using Eq.~\eqref{eq_nonconservative_work_rate} and Eq.~\eqref{eq_external_currents_tw}, $\dot{w}_{\text{nc}}$ becomes
\begin{equation}
\dot{w}_{\text{nc}} = -(\mu_{\alpha_y}-\mu^{\text{ref}}_{\alpha_y})\mathbb S^{\alpha_y}_\rho j^\rho + \nabla\cdot J^{\mathbbm g_Y}+T\dot{\sigma}_{\text{diff}}^{Y}
\label{eq_nc_work_rate_trav_wave}
\end{equation}
The first term on the r.h.s., $-(\mu_{\alpha_y}-\mu^{\text{ref}}_{\alpha_y})\mathbb S^{\alpha_y}_\rho j^\rho$, is the energetic cost for balancing with the chemostats the concentration variations of the $Y$ species due to the chemical reactions. On the other hand, the contribution to the free energy flow $\boldsymbol J^{\mathbbm g_Y}$ and to the diffusion entropy production rate $T\dot{\sigma}_{\text{diff}}^{Y}$ resulting from the controlled species $\alpha_y$ take into account the amount of work needed to prevent modifications to the pattern profiles $Z^{\alpha_y}(\boldsymbol r)$ because of diffusion.

Finally, with the aid of Eqs.~\eqref{eq_time_der_local_free_energy},~\eqref{eq_time_der_local_free_energy_trav_wave} and~\eqref{eq_nc_work_rate_trav_wave}, we find that the entropy production rate reads 
\begin{equation}
T\dot{\sigma}=-\partial_t\mathbbm g - \nabla\cdot J^{\mathbbm g_X} - (\mu_{\alpha_y}-\mu^{\text{ref}}_{\alpha_y})\mathbb S^{\alpha_y}_\rho j^\rho +T\dot{\sigma}_{\text{diff}}^{Y}
\label{eq_entropy_production_rate_trav_wave}
\end{equation}
This general result highlights explicitly which phenomena are responsible for the dissipation: i) a drift contribution due to the propagation of the waves $(\mu_{\alpha_x}-\mu^{\text{ref}}_{\alpha_x})\boldsymbol c\cdot\nabla Z^{\alpha_x}=-\partial_t\mathbbm g$, ii) the free energy flow due to the wave species $\boldsymbol J^{\mathbbm g_X} = (\mu_{\alpha_x}-\mu_{\alpha_x}^{\text{ref}})\boldsymbol J^{\alpha_x}$, iii) the reaction consumption/production of the chemostatted species sustaining the wave dynamics $ (\mu_{\alpha_y}-\mu^{\text{ref}}_{\alpha_y})\mathbb S^{\alpha_y}_\rho j^\rho$ and  iv) the diffusion of the controlled species $T\dot{\sigma}_{\text{diff}}^{Y}$. The first two contributions take into account the dynamics of the waves, while the others represent the dissipation due to the chemostatting of the controlled species. Equation~\eqref{eq_entropy_production_rate_trav_wave} constitutes the second major result of this paper.

The above expressions of the thermodynamic quantities explicitly account for every process occurring while waves propagate. Furthermore, depending on the physical properties of the system of interest, they can be effectively simplified. For example, if the controlled species were homogeneously distributed $Z^{\sigma_y}(\boldsymbol r)=\overline{Z}^{\sigma_y}$ $\forall\boldsymbol r\in V$, no work is required to prevent changes in the controlled species concentrations because of the diffusion. In this case, the nonconservative work rate simplifies to $\dot{w}_{\text{nc}}= -(\mu_{\alpha_y}-\mu^{\text{ref}}_{\alpha_y})\mathbb S^{\alpha_y}_\rho j^\rho$, while the diffusion entropy production rate $T\dot{\sigma}^{Y}_{\text{diff}}$ vanishes. In case the system is detailed balanced, no nonconservative forces are applied, namely $\mu_{\alpha_y}=\mu^{\text{ref}}_{\alpha_y}$, and the dissipation is only due to the propagation of the waves: $T\dot{\sigma}$ depends on the drift and the flow contribution, $T\dot{\sigma}=(\mu_{\alpha_x}-\mu^{\text{ref}}_{\alpha_x})\boldsymbol c\cdot\nabla Z^{\alpha_x} - \nabla\cdot J^{\mathbbm g_X} $. In this case, the global dissipation is solely caused by the transport mechanism of the wave dynamics  $T\dot{\Sigma}=\int_V\mathrm d\boldsymbol r\text{ }(\mu_{\alpha_x}-\mu^{\text{ref}}_{\alpha_x})\boldsymbol c\cdot\nabla Z^{\alpha_x}$.


\section{Model Systems\label{sec_ms}}
We apply here the thermodynamic description developed in the previous section to study the properties of the traveling waves emerging in two model systems. The first model, the Fisher-Kolmogorov equation, is detailed balanced, and the relaxation towards equilibrium occurs via a step-like wave. On the other hand, for the Brusselator model two oscillating traveling waves are maintained by nonconservative forces.

\subsection{Fisher-Kolmogorov equation\label{subsec_fisher}}
A minimal RDS displaying traveling wave solutions is  the so-called Fisher-Kolmogorov equation in one infinite spatial dimension~\cite{Murray2002}. It can be used to describe an autocatalytic reaction between an internal species \ch{X} and a controlled species \ch{Y} according to the chemical equation \ch{X + Y <=>[ $k_{+1}$ ][ $k_{-1}$ ] 2 X}. The system is detailed balanced (no nonconservative forces) since one single species is chemostatted. The reaction-diffusion equation 
\begin{equation}
\partial_t x=  k_{+1}xy- k_{-1}x^2 +D\partial_r^2x,
\end{equation}
with $y$ the homogeneous concentration of the controlled species and $D$ the diffusion coefficient of the internal species, specifies the dynamics of the concentration $Z^{\alpha_x}(r,t)=x(r,t)$ of the internal species. Waves solutions $x(r,t)=x(r-ct)$ emerge for propagation velocities $c$ greater than or equal to a critical value $c\geq \overline{c}=2\sqrt{Dk_{+1}y}$, and they are all characterized by a step-like profile which formally means i) $\lim_{\tilde r\to+\infty} x(\tilde r)=0$, ii) $\lim_{\tilde r\to-\infty} x(\tilde r)=x_{\text{eq}}$ and iii) $\partial_{\tilde r} x<0$ $\forall\tilde{r}$. Note that $\tilde r$ denotes the single spatial coordinate in the frame of reference of the traveling wave, $\tilde r= r -c t$, while $x_{\text{eq}}$ labels the equilibrium concentration of the internal species, $x_{\text{eq}}=k_{+1}y/k_{-1}$, corresponding to the amplitude of the wave. According to these general properties, the thermodynamic quantities of the global system, namely the  entropy production rate $T\dot{\Sigma}$ and the time derivative of the free energy $\mathrm d_t\mathbbm G$ (obtained by integrating the expressions in Eq.~\eqref{eq_time_der_local_free_energy_trav_wave} and~\eqref{eq_entropy_production_rate_trav_wave} over the volume $V$), can be directly related to the features of the traveling wave, namely the amplitude $x_\text{eq}$ and the propagation velocity $c$. Indeed, by considering the infinite volume limit, one obtains
\begin{equation}
T\dot{\Sigma} = -\mathrm d_t\mathbbm G = RT cx_\text{eq}
\end{equation}
establishing a linear dependence between the thermodynamic quantities and the dynamical ones. The above equation, valid for every traveling wave solution, proves that the higher the amplitude and/or the velocity, the greater the dissipation of the global system during the propagation.

Particular traveling wave solutions have to be taken into account if one wants to investigate the local profile of the thermodynamic quantities. For the Fisher-Kolmogoroff equation, both asymptotic solutions and one particular analytical solution are available~\cite{Murray2002}. Since the traveling waves share the same general behavior, we consider the analytical solution 
\begin{equation}
x(\tilde r)=\frac{x_{\text{eq}}}{\left(1+(\sqrt{2}-1)e^{\tilde r\sqrt{k_{+1}y/6D}}\right)^2}
\label{eq_FK_anal_trav_eq}
\end{equation}
with propagation velocity $c=(5/\sqrt{6}) \sqrt{Dk_{+1}y}$, to calculate the  entropy production rate $T\dot{\sigma}$, the free energy flow $\partial_r J^{\mathbbm g}$ and the time derivative of the free energy $\partial_t\mathbbm g$ according to their definitions given in the Eqs.~\eqref{eq_entropy_production_rate_trav_wave},~\eqref{eq_free_energy_diffusion} and~\eqref{eq_time_der_local_free_energy_trav_wave}.
\begin{figure}[t]\centering
\includegraphics[width=1.\columnwidth]{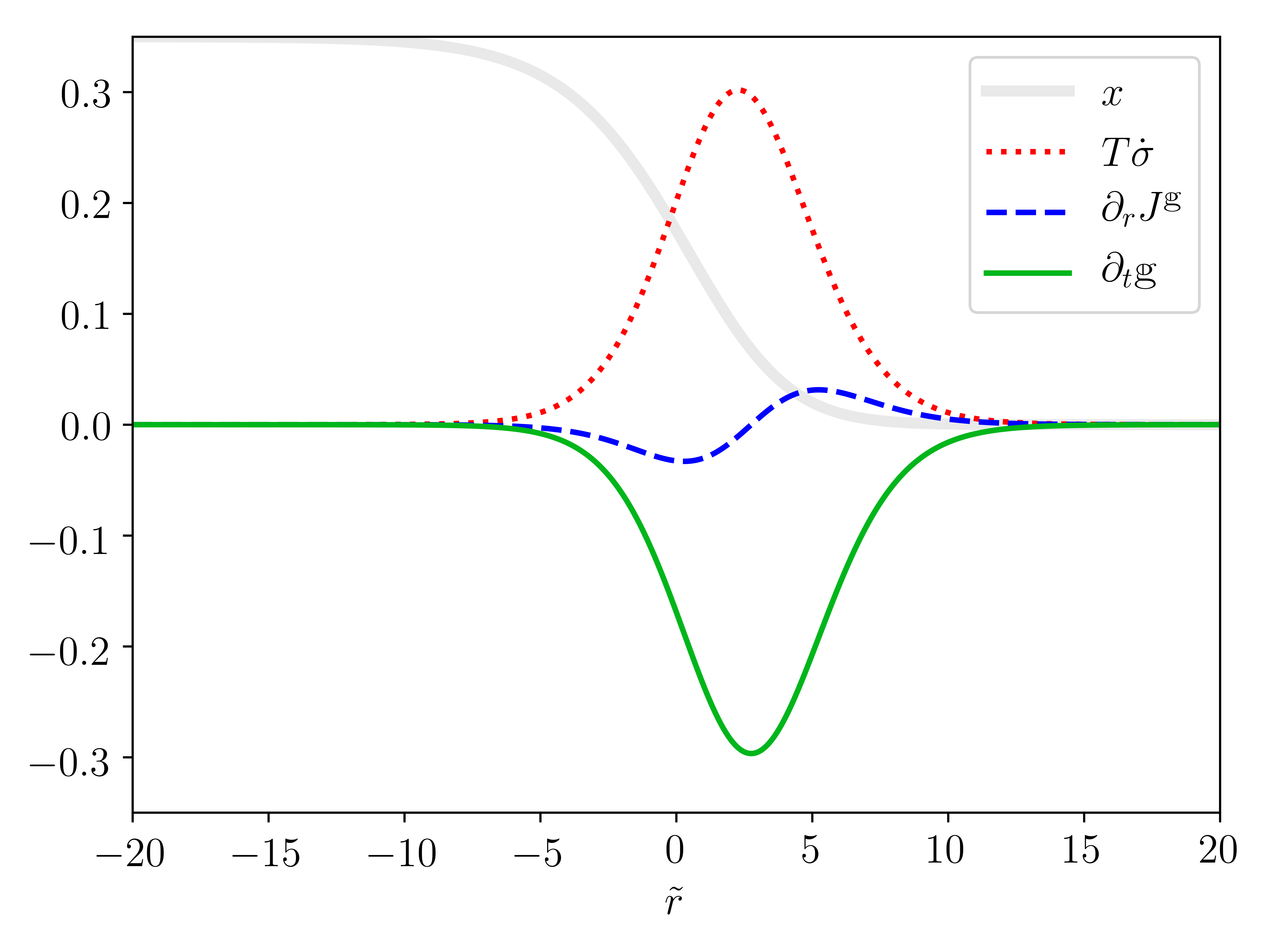}
\caption{Thermodynamic quantities (entropy production rate $T\dot{\sigma}$, free energy flow $\partial_rJ^{\mathbbm g}$ and time derivative of the free energy $\partial_t{\mathbbm g}$) as a function of the comoving coordinate $\tilde r = r-ct$ for the wave solution of the Fisher-Kolmogoroff equation specified in Eq.~\eqref{eq_FK_anal_trav_eq}. We use $RTk_{+1}x_{\text{eq}}y$ and $\sqrt{D/k_{+1}y}$ as units of measure for the thermodynamic quantities and the spatial coordinate, respectively. The profile of the traveling wave $x(\tilde r)$ is scaled by a numerical factor 0.35 to fit in the figure.}
\label{fig_FKeq}
\end{figure}
By inspecting Fig.~\ref{fig_FKeq} where these thermodynamic quantities are displayed, two main features can be noticed. First of all, the variation of the thermodynamic quantities occurs around the wavefront of the concentration field as one would expect. Secondly, the  entropy production rate is larger than the absolute value of the free energy flow, $T\dot{\sigma}>|\partial_rJ^{\mathbbm g}|$, granting that the free energy continuously approaches its equilibrium value even locally, i.e., $\partial_t\mathbbm g<0$ $\forall \tilde r$. In other words, the system is relaxing continuously towards equilibrium for every point of the traveling wave profile. In this model system, the propagation dynamics of the traveling wave does not require to be sustained by nonconservative forces and, consequently, it results to be particularly effective from this specific thermodynamic point of view.

\subsection{Brusselator model\label{subsec_brusselator}}
A simple nondetailed balanced system where oscillating traveling waves emerge is the  Brusselator model in one spatial dimension~\cite{Prigogine1968}. It describes the transformation of two internal species, an activator \ch{X_1} and an inhibitor \ch{X_2}, according to the chemical equations
\begin{align*}
\ch{Y_1 &<=>>[ $k_{+1}$ ][ $k_{-1}$ ] X_1}\\
\ch{X_1 + Y_2 &<=>>[ $k_{+2}$ ][ $k_{-2}$ ] X_2 + Y_3}\\
\ch{2 X_1 + X_2 &<=>>[ $k_{+3}$ ][ $k_{-3}$ ] 3 X_1}\\
\ch{ X_1 &<=>>[ $k_{+4}$ ][ $k_{-4}$ ] Y_4}
\end{align*}
which involve four chemostatted species \ch{Y_1}, \ch{Y_2}, \ch{Y_3}, \ch{Y_4}. We choose \ch{Y_1} and \ch{Y_2} as the reference chemostatted species breaking the two conservation laws and, consequently, determining the equilibrium condition~\cite{Falasco2018a}. The reaction-diffusion equation for the evolution in time and space $r\in V=[0,L]$ of the concentrations $Z^{\alpha_{x_1}}(r,t)=x_1(r,t)$ and $Z^{\alpha_{x_2}}(r,t)=x_2(r,t)$ of the two internal species is specified as
\begin{widetext}
\begin{align}
\begin{split}
&\partial_t x_1= k_{+1}y_1-k_{-1}x_1-k_{+2}x_1y_2+k_{-2}x_2y_3+k_{+3}x_1^2x_2-k_{-3}x_1^3-k_{+4}x_1+k_{-4}y_4+D_1\partial_r^2x_1\\
&\partial_t x_2 = k_{+2}x_1y_2-k_{-2}x_2y_3-k_{+3}x_1^2x_2+k_{-3}x_1^3+D_2\partial_r^2x_2
\end{split}
\label{eq_brusselator_dynamical_system}
\end{align}
\end{widetext}
with $y_1, \dots, y_4$ the homogeneous concentrations of the controlled species, and $D_1$ and $D_2$ the diffusion coefficients of the internal species. Equation~\eqref{eq_brusselator_dynamical_system} admits a uniform steady-state solution $(x_1^{\text{ss}}, x_2^{\text{ss}})$ that becomes unstable if the concentration $y_2$ of the controlled species \ch{Y_2}, used here as a bifurcation parameter, exceeds a critical value, $y_2>\overline{y}_2$. Depending on the critical point, different phenomena emerge including chemical oscillations, Turing patterns, and oscillating traveling waves~\cite{Auchmuty1975, Herschkowitz-Kaufman1975, Auchmuty1976}. The bifurcation point leading to oscillating traveling waves with velocity $c$ and wavenumber $\nu=2\pi n/L$ (with $n\in\mathbb N$ the number of oscillations in the volume $V$ with periodic boundary conditions) is determined by the concentration $\overline y_2$ such that the matrix representing the linearized dynamical system~\eqref{eq_brusselator_dynamical_system} around the steady-state acquires the pure imaginary eigenvalues $\pm\imath\nu\overline{c}$ (with $\overline{c}$ the critical velocity).

In Ref.~\cite{Auchmuty1976}, J. F. G. Auchmuty and G. Nicolis employed a perturbation expansion near the onset of instability, $y_2\simeq\overline y_2$, for identifying the oscillating wave solutions for the concentration $x_1(r,t)$ and $x_2(r,t)$ around their steady-state values in the case of irreversible chemical reactions.  We use here the same analysis (summarized in Appendix~\ref{app_Bruss}) for the reversible model by assuming that the backward reaction currents are almost negligible $j^{-\rho}\simeq 0$ $\forall\rho\in\mathcal R$. This means that the kinetic constants of the forward reactions are greater than the ones of the backward reactions $k_{+\rho}\gg k_{-\rho}$, and the concentrations $y_3$ and $y_4$ are significantly smaller than their equilibrium values, $y_{3,\text{eq}}=k_{+2}k_{+3}y_2/k_{-2}k_{-3}$ and $y_{4,\text{eq}}=k_{+1}k_{+4}y_1/k_{-1}k_{-4}$. In this case,  the existence of traveling wave solutions of Eq.~\eqref{eq_brusselator_dynamical_system} with wavenumber $\nu=2\pi n/ L$ is constrained by the following condition
\begin{equation}
\left(D_2 \nu^2\right)^2-(D_1-D_2)\frac{k_{+1}^2k_{+3}}{k_{+4}^2}(y_1\nu)^2-\frac{k_{+1}^2k_{+3}}{k_{+4}}(y_1)^2\leq0
\label{eq_brusselator_real_velocity}
\end{equation}
granting the existence of the propagation velocity $c$. The corresponding critical concentration is specified as
\begin{equation}
\overline y_2(\nu)=\frac{k_{+4}}{k_{+2}}+\frac{k_{+1}^2k_{+3}}{k_{+2}k_{+4}^2}(y_1)^2+\frac{(D_1+D_2)}{k_{+2}}\nu^2.
\label{eq_brusselator_critical_concentration}
\end{equation}
Equation~\eqref{eq_brusselator_real_velocity} and~\eqref{eq_brusselator_critical_concentration} mean that there is only a finite set of wavenumbers $\nu$  and corresponding critical concentrations $\overline{y}_2(\nu)$ such that the steady state becomes unstable with respect to traveling wave solutions.

Similarly to what has been done for the Fisher-Kolmogorov equation discussed in Subs.~\ref{subsec_fisher}, we establish a correspondence between a global thermodynamic quantity of the Brusselator model and the features of the traveling waves without considering specific solutions. In particular, the global nonconservative work rate $\dot{W}_{\text{nc}}$, i.e., the integral over the volume $V$ of the work rate of Eq.~\eqref{eq_nc_work_rate_trav_wave}, $\dot{W}_{\text{nc}}=-\int_0^L\mathrm dr\text{ }(\mu_{\alpha_y}-\mu^{\text{ref}}_{\alpha_y})\mathbb S^{\alpha_y}_\rho j^\rho$, can be expressed as a function of the wave wavenumber $\nu$. Indeed, by taking into account that i) the chemical potentials of the species \ch{Y_1} and \ch{Y_2} correspond to their reference potentials, $\mu_{\alpha_{y_1}}=\mu^{\text{ref}}_{\alpha_{y_1}}$ and $\mu_{\alpha_{y_2}}=\mu^{\text{ref}}_{\alpha_{y_2}}$, ii) the reaction currents for $\rho=3,\text{ }4$ can be specified as $j^{3}\simeq k_{+2}x_1y_2$ and $j^{4}\simeq k_{+4}x_1$, iii) the oscillation of each traveling wave is centered around the corresponding steady-state concentration $(x_1^{\text{ss}}, x_2^{\text{ss}})\simeq((k_{+1}y_1/k_{+4}), (k_{+2}y_2/k_{+3}x_1^{\text{ss}}))$, the nonconservative work rate reads
\begin{equation}
\dot{W}_{\text{nc}}=LRT k_{+1}y_1\left[\ln\left(\frac{y_{3,\text{eq}}}{y_3}\right)\frac{k_{+2}}{k_{+4}}{y}_2+\ln\left(\frac{y_{4,\text{eq}}}{y_4}\right)\right]
\label{eq_brusselator_global_nc_work_rate}.
\end{equation}
The concentration $y_2$ can be approximated with its critical value $\overline{y}_2(\nu)$ defined in Eq.~\eqref{eq_brusselator_critical_concentration} if wave solutions near the onset of the instability are considered. Since 
$y_{3,\text{eq}}\gg y_{3}$, the global nonconservative work rate is a monotonically increasing function of the wavenumber:
\begin{equation}
\mathrm d_\nu \dot{W}_{\text{nc}}=LRT k_{+1}y_1\ln\left(\frac{y_{3,\text{eq}}}{y_3}\right)\frac{2(D_1+D_2)}{k_{+4}}\nu>0.
\end{equation}
This means that the energetic cost of sustaining oscillating traveling waves for the Brusselator model increases with the wavenumber $\nu$.

\begin{figure}[t]\centering
\includegraphics[width=1.\columnwidth]{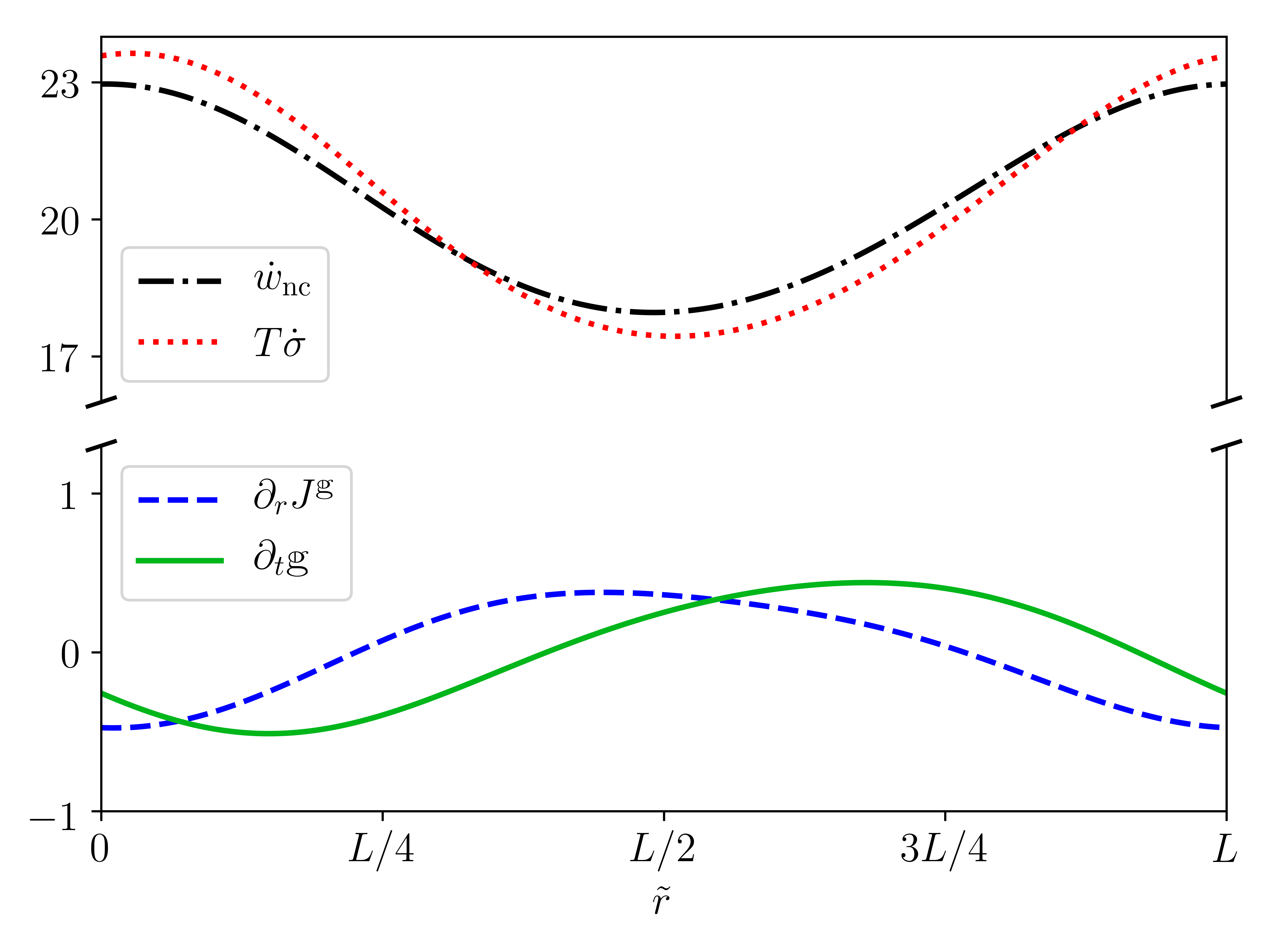}
\caption{Thermodynamic quantities (nonconservative work rate $\dot{w}_{\text{nc}}$, entropy production rate $T\dot{\sigma}$, free energy flow $\partial_rJ^{\mathbbm g}$ and time derivative of the free energy $\partial_t{\mathbbm g}$) as a function of the comoving coordinate $\tilde r = r-ct$ for the oscillating wave solutions of Brusselator model with $\nu=2\pi/L$. We use $RT/k_{+2}$ and $\sqrt{D_1/k_{+1}}$ as units of measure for the thermodynamic quantities and the spatial coordinate, respectively.  }
\label{fig_Bruss}
\end{figure}

We consider now the specific traveling wave solutions,
\begin{equation}
\small
\begin{split}
& x_1(\tilde r)= x_1^{\text{ss}} + \epsilon \cos(\nu\tilde r) + 2 \epsilon^2\phi_1\cos(2\nu\tilde r+ \theta_1) \\
& x_2 (\tilde r)=x_2^{\text{ss}} - \epsilon\overline y_2 \phi \cos(\nu\tilde r+\theta) + \epsilon^2(d+ 2\phi_2\cos(2\nu\tilde r+ \theta_2))
\end{split}
\label{eq_brusselator_traveling_waves}
\end{equation}
whose derivation is discussed in Appendix~\ref{app_Bruss}, to study the local profile of the thermodynamic quantities defined in Eqs.~\eqref{eq_time_der_local_free_energy_trav_wave}, \eqref{eq_nc_work_rate_trav_wave}, \eqref{eq_entropy_production_rate_trav_wave} and~\eqref{eq_free_energy_diffusion}, and displayed in Fig.~\ref{fig_Bruss} for the case with $n=1$. The particular values of the physical quantities employed for the plot, like the concentration of the chemostatted species, are reported in Appendix~\ref{app_Bruss}.
First of all, one can verify that all the thermodynamic quantities share the same period as the traveling waves. Secondly, there is a significant difference between the nonconservative work rate and the entropy production rate, on the one hand,  and the time derivative of the free energy and the divergence of its flow, on the other hand. The nonconservative work spent to sustain the propagation of the waves is mainly dissipated, $\dot{w}_{\text{nc}}\sim T\dot{\sigma}$, resulting in a limited variation of the free energy in time $\dot{w}_{\text{nc}}\sim T\dot{\sigma} \gg| \partial_t\mathbbm g|$. This behavior is caused by the need of keeping the system far from equilibrium to allow the emergence of oscillating traveling waves. From this thermodynamic point of view, the propagation of oscillating traveling waves in the Brusselator model is energetically very expensive: it requires that a lot of work is performed.

A similar general behavior is observed if one considers traveling waves emerging under different conditions, such as different concentrations of the controlled species and different wavenumber. In this regard, an interesting difference is the increase of the global nonconservative work rate with the wavenumbers according to Eq.~\eqref{eq_brusselator_global_nc_work_rate}. A similar trend can also be observed for the amplitude of the oscillating nonconservative work rate, as well as the amplitude of the free energy.

\section{Conclusions\label{sec_conc}}
In this work, we established a local thermodynamic theory of reaction-diffusion systems valid arbitrarily far from equilibrium. We then used it to investigate the propagation of chemical waves and its energetic cost. To this aim, the identification of the proper thermodynamic potential as well as establishing its balance equations have been the fundamental steps. We showed a clear correspondence between the thermodynamic quantities and the dynamical processes occurring while chemical waves propagate, both in general and for two model systems. The global entropy production rate of the Fisher- Kolmogorov waves is linearly dependent on the velocity and the wave amplitude. The global nonconservative work rate is a monotonically increasing function of the wavenumber for the periodic waves of the Brusselator model. The relationship between the features of the concentration profiles of the chemical waves and the thermodynamic quantities has also been analyzed in these models at the local level. 


We focused here on the propagation dynamics of chemical waves, but our thermodynamic description could be used to study any other reaction-diffusion processes. A direct application might be the characterization of the energetic cost of creating wave patterns from generic initial conditions and of the response to perturbations in model systems like those studied in this work. Furthermore, we think that our approach represents the proper framework to study the correspondence between the semantic information codified in complex chemical patterns like chemical waves and the syntactic information. The semantic information represents the information content relevant for a specific system which depends on how the patterns are interpreted by for instance a receptor while the syntactic information represents the whole physical information content~\cite{Kolchinsky2018}. These topics are becoming more important nowadays since they play a crucial role in the development of new ways of transferring and computing information with chemical reactions~\cite{Rossi2008,Padirac2013,Tayar2015,Adamatzky2005}. They should also shed light on how biosystems perform these operations with high efficiency~\cite{Sartori2014, Barato2014, Ito2015}.


Finally, our work is based on a deterministic description of the reaction-diffusion dynamics, which emerges in the large size limit of a microscopic description in terms of stochastically reacting and diffusing chemicals. A thermodynamic at that level is necessary to describe intracellular signaling because of the limited number of molecules. For instance, the importance of stochasticity in intracellular calcium signaling is still debated~\cite{Falcke2003, Skupin2010}.


\section{Acknowledgments}
This research was funded by the European Research Council project NanoThermo (ERC-2015-CoG Agreement No.~681456).


\appendix
\section{Reference Chemical Potentials\label{app_ref_chem_pot}}
We examine here the relations between the reference chemical potentials of different species $\mu^{\text{ref}}_{\alpha}$ for open RDSs. Note that the classification of the species $\mathcal S$ as internal species $X$ or controlled species breaking the broken conservation laws $Y_b$ or other controlled species $Y_p$ is particularly important in this framework. As already mentioned in the main text, the reference chemical potentials are the values of the chemical potentials at the equilibrium that would be reached if the system were detailed balanced. For this reason and the local detailed balanced condition $k_{+\rho}/k_{-\rho}=\prod_\alpha (Z^{\alpha}_{\text{eq}})^{\mathbb S^{\alpha}_{\rho}}=\exp(-\mathbb S^{\alpha}_\rho\mu_\alpha^{\circ}/RT)$, the set of reference chemical potentials is a left null vector of the stoichiometric matrix,
\begin{equation}
\mu^{\text{ref}}_{\alpha}\mathbb S^{\alpha}_\rho=0,\text{ }\forall\rho\in\mathcal R.
\label{eq_vanish_gibbs_free_energy_rct}
\end{equation}
This means that the reference chemical potentials $\mu^{\text{ref}}_{\alpha}$ can be written as a linear combination of the closed system conservation laws $l^\lambda_\alpha$,
\begin{equation}
\mu^{\text{ref}}_{\alpha} = \psi_\lambda l^{\lambda}_\alpha
\label{eq_mu_as_linear_comb_l}
\end{equation}
where $\psi_\lambda$ are real coefficients. Because of Eq.~\eqref{eq_mu_as_linear_comb_l}, the number of independent reference chemical potentials is equal to the number of conservation laws $|\lambda|$ of the closed RDS. The other chemical potentials are determined according to the constraints of Eq.~\eqref{eq_vanish_gibbs_free_energy_rct}.

First of all, we show that the reference chemical potentials of the chemostatted species which break all the broken conservation laws $\mu^{\text{ref}}_{\alpha_{y_b}}$ set the value of the reference chemical potentials of the other controlled species $\mu^{\text{ref}}_{\alpha_{y_p}}$. By definition of unbroken conservation laws, i.e., $l^{\lambda_u}_{\alpha_x}\mathbb S^{\alpha_x}_{\rho}=0$ $\forall\rho\in\mathcal R$ and $l^{\lambda_u}_{\alpha_y}=0$, the reference chemical potentials $\mu^{\text{ref}}_{\alpha_{y_b}}$ (resp. $\mu^{\text{ref}}_{\alpha_{y_p}}$) are defined as a linear combination of only the broken conservation laws, $\mu^{\text{ref}}_{\alpha_{y_b}}=\psi_{\lambda_b}l^{\lambda_b}_{\alpha_{y_b}}$ (resp. $\mu^{\text{ref}}_{\alpha_{y_p}}=\psi_{\lambda_b}l^{\lambda_b}_{\alpha_{y_p}}$). The matrix whose entries are $l^{\lambda_b}_{\alpha_{y_b}}$ is square and nonsingular so that it can be inverted. The elements of the inverse matrix are denoted by $\hat{l}_{\lambda_b}^{\alpha_{y_b}}$, and the reference chemical potentials $\mu^{\text{ref}}_{\alpha_{y_p}}$ can now be specified as
\begin{equation}
\mu^{\text{ref}}_{\alpha_{y_p}}=\mu^{\text{ref}}_{\alpha_{y_b}}\hat{l}_{\lambda_b}^{\alpha_{y_b}}{l}^{\lambda_b}_{\alpha_{y_p}},
\end{equation}
since $\psi_{\lambda_b}=\mu^{\text{ref}}_{\alpha_{y_b}}\hat{l}_{\lambda_b}^{\alpha_{y_b}}$.

Secondly, we show that the value of reference chemical potentials for the internal species $\mu^{\text{ref}}_{\alpha_x}$ depends on the reference chemical potentials $\mu^{\text{ref}}_{\alpha_{y_b}}$ and the unbroken conserved quantities $L^{\lambda_u}=\int_V\mathrm d\boldsymbol r\text{ }l^{\lambda_u}_{\alpha_x}Z^{\alpha_x}$. Because of $\psi_{\lambda_b}=\mu^{\text{ref}}_{\alpha_{y_b}}\hat{l}_{\lambda_b}^{\alpha_{y_b}}$, 
\begin{equation}
\mu^{\text{ref}}_{\alpha_{x}}=\psi_{\lambda_u}l^{\lambda_u}_{\alpha_x}+\mu^{\text{ref}}_{\alpha_{y_b}}\hat{l}_{\lambda_b}^{\alpha_{y_b}}{l}^{\lambda_b}_{\alpha_{x}},
\end{equation}
with the coefficients $\psi_{\lambda_u}$ to be determined by employing the unbroken conserved quantities evaluated for the equilibrium concentrations $L^{\lambda_u}=Vl^{\lambda_u}_{\alpha_x}Z^{\alpha_x}_{\text{eq}}$. Indeed, the conserved quantities can be written as 
\begin{equation}
\small
L^{\lambda_u}=V\sum_{\alpha_x\in X}l^{\lambda_u}_{\alpha_x}\exp\left\{(\psi_{\lambda_u'}l^{\lambda_u'}_{\alpha_x}+\mu^{\text{ref}}_{\alpha_{y_b}}\hat{l}_{\lambda_b}^{\alpha_{y_b}}{l}^{\lambda_b}_{\alpha_{x}}-\mu_{\alpha_x}^{\circ})/RT\right\}\label{eq_not_inv}
\end{equation}
establishing $|\lambda_u|$ nonlinear constrains for the $|\lambda_u|$ coefficients $\psi_{\lambda_u}$. By solving the above system of equations for the coefficients $\psi_{\lambda_u}$, one obtains $\psi_{\lambda_u}$ as a function of $\{L^{\lambda_u}\}$ and $\{\mu^{\text{ref}}_{\alpha_{y_b}}\}$. Unfortunately, explicit expressions  cannot be obtained from Eq.~\eqref{eq_not_inv}. If there are no conserved quantities, $\mu_{\alpha_{x}}^{\text{ref}}=\mu_{\alpha_{y_b}}^{\text{ref}}\hat{l}^{\alpha_{y_b}}_{\lambda_b}l^{\lambda_b}_{\alpha_x}$. In conclusion, the reference chemical potentials $\mu^{\text{ref}}_{\alpha_{y_b}}$ and the conserved quantities $L^{\lambda_u}$ set the values of all the other reference chemical potentials $\mu^{\text{ref}}_{\alpha_{x}}$, $\mu^{\text{ref}}_{\alpha_{y_p}}$.

\section{Traveling Waves in the Brusselator Model\label{app_Bruss}}
We summarize here the derivation of oscillating traveling wave solutions of the Brusselator model  in one spatial dimension $r\in[0,L]$ with irreversible chemical reactions proposed by J. F. G. Auchmuty and G. Nicolis~\cite{Auchmuty1976}. Since the backward reaction currents are neglected, the reaction-diffusion equation~\eqref{eq_brusselator_dynamical_system} reads, in vector form,
\begin{align}
\small
\partial_t
\begin{pmatrix}
x_1\\
x_2
\end{pmatrix} = 
\begin{pmatrix}
 k_{+1}y_1-k_{+2}x_1y_2+k_{+3}x_1^2x_2-k_{+4}x_1+D_1\partial_r^2x_1\\
k_{+2}x_1y_2-k_{+3}x_1^2x_2+D_2\partial_r^2x_2
\end{pmatrix}.
\label{eq_IRR_brusselator_dynamical_system}
\end{align}
Oscillating traveling wave solutions of Eq.~\eqref{eq_IRR_brusselator_dynamical_system} are periodic functions of the only comoving coordinate: $x_1(r,t)=x_1(\tilde r)= x_1(\tilde r\pm L)$ and $x_2(r,t)=x_2(\tilde r)= x_2(\tilde r\pm L)$, with $\tilde r = r-ct$. In this regard, the steady-state $(x_1^{\text{ss}}, x_2^{\text{ss}})=((k_{+1}y_1/k_{+4}), (k_{+2}y_2/k_{+3}x_1^{\text{ss}}))$ can be interpreted as a trivial traveling wave with null velocity. By considering the linearization of the dynamical system~\eqref{eq_IRR_brusselator_dynamical_system} around the steady-state according to the operator
\begin{align}
\small
\hat{\mathcal J}=
{\begin{pmatrix}
 &k_{+2}y_2-k_{+4}+D_1\partial_r^2 &k_{+1}^2k_{+3}(y_1)^2/k_{+4}^2\\
&-k_{+2}y_2 &-k_{+1}^2k_{+3}(y_1)^2/k_{+4}^2+D_2\partial_r^2
\end{pmatrix}},
\label{eq_IRR_LIN_brusselator_dynamical_system}
\end{align}
one can identify the conditions where new wave solutions may bifurcate from the steady-state. Any such a solution, written as a perturbation around the steady-state $(x_1^{\text{p}},x_2^{\text{p}})=(x_1-x_1^{\text{ss}},x_2 - x_2^{\text{ss}})$, must have the following form in the linear regime
\begin{equation}
\begin{pmatrix}
x_1^{\text{p}}\\
x_2^{\text{p}}
\end{pmatrix}=
\begin{pmatrix}
a_1\\
a_2
\end{pmatrix}e^{\imath\nu\tilde r}
\label{eq_linear_tw_solution_brusselator}
\end{equation}
with $(a_1, a_2)$ still unknown coefficients and $\nu=2\pi n/L$ ($n\in\mathbb N$), because of the periodic boundary conditions. Consequently, $(x_1^{\text{p}}, x_2^{\text{p}})$ of Eq.~\eqref{eq_linear_tw_solution_brusselator} must be an eigenvector of the linear operator $\hat{\mathcal J}$ corresponding to the complex eigenvalue $-\imath\nu c$. To guarantee the existence of such a solution, the matrix representation of the operator $\hat{\mathcal J}$ on the basis element $\exp(\imath\nu\tilde r)$
\begin{align}
\small
\mathcal J= 
\begin{pmatrix}
 &k_{+2}y_2-k_{+4}-D_1\nu^2 &k_{+1}^2k_{+3}(y_1)^2/k_{+4}^2\\
&-k_{+2}y_2 &-k_{+1}^2k_{+3}(y_1)^2/k_{+4}^2-D_2\nu^2
\end{pmatrix}
\label{eq_IRR_LIN_matrix_brusselator}
\end{align}
must have purely complex eigenvalues meaning that its trace vanishes $\mathcal T=0$, while its determinant is positive $\mathcal D>0$. The condition $\mathcal T=0$ determines the critical value $\overline{y}_2$ of the concentration of the chemostatted species \ch{Y_2}, considered here as the bifurcation parameter, which is specified in Eq.~\eqref{eq_brusselator_critical_concentration}. On the other hand, the constraint $\mathcal D>0$ is equivalent to Eq.~\eqref{eq_brusselator_real_velocity} when the critical concentration $\overline{y}_2$ is employed, and it ensures that the value of the critical velocity $\overline{c}(\nu)$, namely  the wave velocity evaluated for $y_2=\overline{y}_2$, is a real number:
\begin{equation}
\small
\overline{c}(\nu)=(D_1-D_2)\frac{k_{+1}^2k_{+3}}{k_{+4}^2}(y_1)^2+\frac{k_{+1}^2k_{+3}}{k_{+4}}\left(\frac{y_1}{\nu}\right)^2-\left(D_2 \nu\right)^2.
\end{equation}

When the concentration $y_2$ exceeds its critical value $\overline{y}_2(\nu)$, the steady-state becomes unstable compared to the traveling waves with wavenumber $\nu$. Waves solutions of the reaction-diffusion equation close to the bifurcation point can be identified by expanding the dynamical system~\eqref{eq_IRR_brusselator_dynamical_system} in powers of a small parameter $\epsilon$ representing the distance from the threshold. To this aim, Eq.~\eqref{eq_IRR_brusselator_dynamical_system} can be written in such a way that i) the difference between the actual concentration $y_2$ (resp. velocity $c$) and the critical one $\overline{y}_2(\nu)$ (resp. $\overline c(\nu)$) is made explicit, and ii) the linear contributions to the dynamics are gathered together and separated from the nonlinear one. In this way, Eq.~\eqref{eq_IRR_brusselator_dynamical_system} reads
\begin{widetext}
\begin{align}
\small
-(c-\overline c)
\begin{pmatrix}
x_1\\
x_2
\end{pmatrix} = 
(\hat{\overline{\mathcal J}}+\overline c )
\begin{pmatrix}
x_1\\
x_2
\end{pmatrix}
+ k_2(y_2-\overline{y}_2)
\begin{pmatrix}
x_1\\
-x_1
\end{pmatrix}
+
\begin{pmatrix}
h(x_1, x_2)\\
-h(x_1,x_2)
\end{pmatrix}
\label{eq_IRR_brusselator_dynamical_system_BIS}
\end{align}
\end{widetext}
where $\hat{\overline{\mathcal J}}$ is the operator $\hat{{\mathcal J}}$ evaluated at the critical concentration $\overline{y}_2(\nu)$, while $h(x_1, x_2)$ takes into account the nonlinear dynamics. We shall now expand every term in Eq.~\eqref{eq_IRR_brusselator_dynamical_system_BIS} in powers of $\epsilon$
\begin{equation}
\small
\begin{gathered}
\begin{pmatrix}
x_1\\
x_2
\end{pmatrix}= \sum_{q=1}^{+\infty}\epsilon^q\begin{pmatrix}
x_1^{(q)}\\
x_2^{(q)}
\end{pmatrix}, \\
 y_2-\overline{y}_2(\nu)=\sum_{q=1}^{+\infty} \epsilon^qy^{(q)}, \\
c-\overline{c}(\nu)=\sum_{q=1}^{+\infty}  \epsilon^qc^{(q)},
\end{gathered}
\end{equation}
and then iteratively solve each equation obtained by gathering equal powers of $\epsilon$. Following this perturbation procedure, one derives the traveling wave solutions with wavenumber $\nu$ of the Brusselator model. The details of the perturbation expansion in power of $\epsilon$ are discussed in Ref.~\cite{Auchmuty1976}, and they are not reported here.

The particular traveling wave solution in Eq.~\eqref{eq_brusselator_traveling_waves}, as well as any other quantities mentioned in the following, is specified according to the set of unit measures reported in the Table~\ref{tab_unit_measures}.
\begin{table}\centering
\caption{\label{tab_unit_measures} Set of units measures adopted for the Brusselator model}
\begin{tabular}{cc}
\toprule
Physical quantity & Unit measures\\
\colrule
time & $1/k_{+1}$   \\
space & $\sqrt{D_1/k_{+1}}$  \\
concentration & $k_{+1}/k_{+2}$\\
\botrule
\end{tabular}
\end{table}
It has been obtained under the hypothesis of kinetic constants $k_{+\rho}=1$ and stopping the perturbation expansion at the second order. The explicit expressions of the coefficients  $\epsilon$, $\phi$, $\phi_1$, $\phi_2$, $\theta$, $\theta_1$, $\theta_2$ can be found in Ref.~\cite{Auchmuty1976}. For the plot in Fig.~\ref{fig_Bruss}, we adopt the following numerical values of the physical quantities: $y_1=0.5$, $y_2=1.01\overline{y}_2$, $D_2=0.05$, $y_3=10$, $y_4=1$ and $k_{-\rho}=10^{-4}$. Similar results to those discussed in the Subs.~\ref{subsec_brusselator} are obtained with different numerical values as long as the concentrations of the chemostatted species \ch{Y_3} and \ch{Y_4} are much smaller than their equilibrium ones.

Finally, it has to be mentioned that our thermodynamic theory is based on the implicit assumption that a reference equilibrium condition is well defined. This means that it can be applied only to the Brusselator model with reversible chemical reactions. On the other hand, we employ the analytical traveling wave solutions of the irreversible Brusselator as an approximate expression of the traveling wave solutions for the case of negligible, but not null, backward reaction currents. This choice appears to be reasonable since the relations between the thermodynamic quantities, like $\partial_t\mathbbm g=-T\dot{\sigma}+\dot{w}_{\text{nc}}-\partial_rJ^{\mathbbm g}$, are still verified.

\bibliography{biblio}

\begin{thebibliography}{40}%
\makeatletter
\providecommand \@ifxundefined [1]{%
 \@ifx{#1\undefined}
}%
\providecommand \@ifnum [1]{%
 \ifnum #1\expandafter \@firstoftwo
 \else \expandafter \@secondoftwo
 \fi
}%
\providecommand \@ifx [1]{%
 \ifx #1\expandafter \@firstoftwo
 \else \expandafter \@secondoftwo
 \fi
}%
\providecommand \natexlab [1]{#1}%
\providecommand \enquote  [1]{``#1''}%
\providecommand \bibnamefont  [1]{#1}%
\providecommand \bibfnamefont [1]{#1}%
\providecommand \citenamefont [1]{#1}%
\providecommand \href@noop [0]{\@secondoftwo}%
\providecommand \href [0]{\begingroup \@sanitize@url \@href}%
\providecommand \@href[1]{\@@startlink{#1}\@@href}%
\providecommand \@@href[1]{\endgroup#1\@@endlink}%
\providecommand \@sanitize@url [0]{\catcode `\\12\catcode `\$12\catcode
  `\&12\catcode `\#12\catcode `\^12\catcode `\_12\catcode `\%12\relax}%
\providecommand \@@startlink[1]{}%
\providecommand \@@endlink[0]{}%
\providecommand \url  [0]{\begingroup\@sanitize@url \@url }%
\providecommand \@url [1]{\endgroup\@href {#1}{\urlprefix }}%
\providecommand \urlprefix  [0]{URL }%
\providecommand \Eprint [0]{\href }%
\providecommand \doibase [0]{http://dx.doi.org/}%
\providecommand \selectlanguage [0]{\@gobble}%
\providecommand \bibinfo  [0]{\@secondoftwo}%
\providecommand \bibfield  [0]{\@secondoftwo}%
\providecommand \translation [1]{[#1]}%
\providecommand \BibitemOpen [0]{}%
\providecommand \bibitemStop [0]{}%
\providecommand \bibitemNoStop [0]{.\EOS\space}%
\providecommand \EOS [0]{\spacefactor3000\relax}%
\providecommand \BibitemShut  [1]{\csname bibitem#1\endcsname}%
\let\auto@bib@innerbib\@empty
\bibitem [{\citenamefont {Murray}(2002)}]{Murray2002}%
  \BibitemOpen
  \bibfield  {author} {\bibinfo {author} {\bibfnamefont {J.~D.}\ \bibnamefont
  {Murray}},\ }\href {\doibase 10.1007/b98868} {\emph {\bibinfo {title}
  {Mathematical Biology I. An Introduction}}},\ \bibinfo {edition} {3rd}\ ed.,\
  \bibinfo {series} {Interdisciplinary Applied Mathematics}, Vol.~\bibinfo
  {volume} {17}\ (\bibinfo  {publisher} {Springer},\ \bibinfo {address} {New
  York},\ \bibinfo {year} {2002})\BibitemShut {NoStop}%
\bibitem [{\citenamefont {Deneke}\ and\ \citenamefont
  {Di~Talia}(2018)}]{Deneke2018}%
  \BibitemOpen
  \bibfield  {author} {\bibinfo {author} {\bibfnamefont {V.~E.}\ \bibnamefont
  {Deneke}}\ and\ \bibinfo {author} {\bibfnamefont {S.}~\bibnamefont
  {Di~Talia}},\ }\href {\doibase 10.1083/jcb.201701158} {\bibfield  {journal}
  {\bibinfo  {journal} {J. Cell Biol.}\ }\textbf {\bibinfo {volume} {217}},\
  \bibinfo {pages} {1193} (\bibinfo {year} {2018})}\BibitemShut {NoStop}%
\bibitem [{\citenamefont {Purvis}\ and\ \citenamefont
  {Lahav}(2013)}]{Purvis2013}%
  \BibitemOpen
  \bibfield  {author} {\bibinfo {author} {\bibfnamefont {J.~E.}\ \bibnamefont
  {Purvis}}\ and\ \bibinfo {author} {\bibfnamefont {G.}~\bibnamefont {Lahav}},\
  }\href {\doibase 10.1016/j.cell.2013.02.005} {\bibfield  {journal} {\bibinfo
  {journal} {Cell}\ }\textbf {\bibinfo {volume} {152}},\ \bibinfo {pages} {945}
  (\bibinfo {year} {2013})}\BibitemShut {NoStop}%
\bibitem [{\citenamefont {Behar}\ and\ \citenamefont
  {Hoffmann}(2010)}]{Behar2010}%
  \BibitemOpen
  \bibfield  {author} {\bibinfo {author} {\bibfnamefont {M.}~\bibnamefont
  {Behar}}\ and\ \bibinfo {author} {\bibfnamefont {A.}~\bibnamefont
  {Hoffmann}},\ }\href {\doibase https://doi.org/10.1016/j.gde.2010.09.007}
  {\bibfield  {journal} {\bibinfo  {journal} {Curr. Opin. Genetics Dev.}\
  }\textbf {\bibinfo {volume} {20}},\ \bibinfo {pages} {684 } (\bibinfo {year}
  {2010})}\BibitemShut {NoStop}%
\bibitem [{\citenamefont {Berridge}\ \emph {et~al.}(2003)\citenamefont
  {Berridge}, \citenamefont {Bootman},\ and\ \citenamefont
  {Roderick}}]{Berridge2003}%
  \BibitemOpen
  \bibfield  {author} {\bibinfo {author} {\bibfnamefont {M.~J.}\ \bibnamefont
  {Berridge}}, \bibinfo {author} {\bibfnamefont {M.~D.}\ \bibnamefont
  {Bootman}}, \ and\ \bibinfo {author} {\bibfnamefont {H.~L.}\ \bibnamefont
  {Roderick}},\ }\href {https://doi.org/10.1038/nrm1155} {\bibfield  {journal}
  {\bibinfo  {journal} {Nat. Rev. Mol. Cell Biol.}\ }\textbf {\bibinfo {volume}
  {4}},\ \bibinfo {pages} {517} (\bibinfo {year} {2003})}\BibitemShut {NoStop}%
\bibitem [{\citenamefont {Berridge}\ \emph {et~al.}(2000)\citenamefont
  {Berridge}, \citenamefont {Lipp},\ and\ \citenamefont
  {Bootman}}]{Berridge2000}%
  \BibitemOpen
  \bibfield  {author} {\bibinfo {author} {\bibfnamefont {M.~J.}\ \bibnamefont
  {Berridge}}, \bibinfo {author} {\bibfnamefont {P.}~\bibnamefont {Lipp}}, \
  and\ \bibinfo {author} {\bibfnamefont {M.~D.}\ \bibnamefont {Bootman}},\
  }\href {https://doi.org/10.1038/35036035} {\bibfield  {journal} {\bibinfo
  {journal} {Nat. Rev. Mol. Cell Biol.}\ }\textbf {\bibinfo {volume} {1}},\
  \bibinfo {pages} {11 } (\bibinfo {year} {2000})}\BibitemShut {NoStop}%
\bibitem [{\citenamefont {Carafoli}\ \emph {et~al.}(2001)\citenamefont
  {Carafoli}, \citenamefont {Santella}, \citenamefont {Branca},\ and\
  \citenamefont {Brini}}]{Carafoli2001}%
  \BibitemOpen
  \bibfield  {author} {\bibinfo {author} {\bibfnamefont {E.}~\bibnamefont
  {Carafoli}}, \bibinfo {author} {\bibfnamefont {L.}~\bibnamefont {Santella}},
  \bibinfo {author} {\bibfnamefont {D.}~\bibnamefont {Branca}}, \ and\ \bibinfo
  {author} {\bibfnamefont {M.}~\bibnamefont {Brini}},\ }\href {\doibase
  10.1080/20014091074183} {\bibfield  {journal} {\bibinfo  {journal} {Crit.
  Rev. Biochem. Mol. Biol.}\ }\textbf {\bibinfo {volume} {36}},\ \bibinfo
  {pages} {107} (\bibinfo {year} {2001})}\BibitemShut {NoStop}%
\bibitem [{\citenamefont {Prigogine}\ and\ \citenamefont
  {Nicolis}(1971)}]{Prigogine1971}%
  \BibitemOpen
  \bibfield  {author} {\bibinfo {author} {\bibfnamefont {I.}~\bibnamefont
  {Prigogine}}\ and\ \bibinfo {author} {\bibfnamefont {G.}~\bibnamefont
  {Nicolis}},\ }\href {\doibase 10.1017/S0033583500000615} {\bibfield
  {journal} {\bibinfo  {journal} {Q. Rev. Biophys.}\ }\textbf {\bibinfo
  {volume} {4}},\ \bibinfo {pages} {107} (\bibinfo {year} {1971})}\BibitemShut
  {NoStop}%
\bibitem [{\citenamefont {Nicolis}\ and\ \citenamefont
  {Prigogine}(1977)}]{Nicolis1977}%
  \BibitemOpen
  \bibfield  {author} {\bibinfo {author} {\bibfnamefont {G.}~\bibnamefont
  {Nicolis}}\ and\ \bibinfo {author} {\bibfnamefont {I.}~\bibnamefont
  {Prigogine}},\ }\href@noop {} {\emph {\bibinfo {title} {Self-organization in
  Nonequilibrium Systems: From Dissipative Structures to Order Through
  Fluctuations}}}\ (\bibinfo  {publisher} {Wiley-Blackwell},\ \bibinfo {year}
  {1977})\BibitemShut {NoStop}%
\bibitem [{\citenamefont {Cross}\ and\ \citenamefont
  {Hohenberg}(1993)}]{Cross1993a}%
  \BibitemOpen
  \bibfield  {author} {\bibinfo {author} {\bibfnamefont {M.~C.}\ \bibnamefont
  {Cross}}\ and\ \bibinfo {author} {\bibfnamefont {P.~C.}\ \bibnamefont
  {Hohenberg}},\ }\href {\doibase 10.1103/RevModPhys.65.851} {\bibfield
  {journal} {\bibinfo  {journal} {Rev. Mod. Phys.}\ }\textbf {\bibinfo {volume}
  {65}},\ \bibinfo {pages} {851} (\bibinfo {year} {1993})}\BibitemShut
  {NoStop}%
\bibitem [{\citenamefont {Ross}(2008)}]{Ross2008}%
  \BibitemOpen
  \bibfield  {author} {\bibinfo {author} {\bibfnamefont {J.}~\bibnamefont
  {Ross}},\ }\href {\doibase 10.1007/978-3-540-74555-6} {\emph {\bibinfo
  {title} {Thermodynamics and Fluctuations far from Equilibrium}}}\ (\bibinfo
  {publisher} {Springer},\ \bibinfo {year} {2008})\BibitemShut {NoStop}%
\bibitem [{\citenamefont {Rao}\ and\ \citenamefont {Esposito}(2016)}]{Rao2016}%
  \BibitemOpen
  \bibfield  {author} {\bibinfo {author} {\bibfnamefont {R.}~\bibnamefont
  {Rao}}\ and\ \bibinfo {author} {\bibfnamefont {M.}~\bibnamefont {Esposito}},\
  }\href {\doibase 10.1103/PhysRevX.6.041064} {\bibfield  {journal} {\bibinfo
  {journal} {Phys. Rev. X}\ }\textbf {\bibinfo {volume} {6}},\ \bibinfo {pages}
  {041064} (\bibinfo {year} {2016})}\BibitemShut {NoStop}%
\bibitem [{\citenamefont {Rao}\ and\ \citenamefont
  {Esposito}(2018{\natexlab{a}})}]{Rao2018b}%
  \BibitemOpen
  \bibfield  {author} {\bibinfo {author} {\bibfnamefont {R.}~\bibnamefont
  {Rao}}\ and\ \bibinfo {author} {\bibfnamefont {M.}~\bibnamefont {Esposito}},\
  }\href {\doibase 10.1063/1.5042253} {\bibfield  {journal} {\bibinfo
  {journal} {J. Chem. Phys.}\ }\textbf {\bibinfo {volume} {149}},\ \bibinfo
  {pages} {245101} (\bibinfo {year} {2018}{\natexlab{a}})}\BibitemShut
  {NoStop}%
\bibitem [{\citenamefont {Falasco}\ \emph {et~al.}(2018)\citenamefont
  {Falasco}, \citenamefont {Rao},\ and\ \citenamefont
  {Esposito}}]{Falasco2018a}%
  \BibitemOpen
  \bibfield  {author} {\bibinfo {author} {\bibfnamefont {G.}~\bibnamefont
  {Falasco}}, \bibinfo {author} {\bibfnamefont {R.}~\bibnamefont {Rao}}, \ and\
  \bibinfo {author} {\bibfnamefont {M.}~\bibnamefont {Esposito}},\ }\href
  {\doibase 10.1103/PhysRevLett.121.108301} {\bibfield  {journal} {\bibinfo
  {journal} {Phys. Rev. Lett.}\ }\textbf {\bibinfo {volume} {121}},\ \bibinfo
  {pages} {108301} (\bibinfo {year} {2018})}\BibitemShut {NoStop}%
\bibitem [{\citenamefont {Jarzynski}(2011)}]{Jarzynski2011}%
  \BibitemOpen
  \bibfield  {author} {\bibinfo {author} {\bibfnamefont {C.}~\bibnamefont
  {Jarzynski}},\ }\href {\doibase 10.1146/annurev-conmatphys-062910-140506}
  {\bibfield  {journal} {\bibinfo  {journal} {Annu. Rev. Condens. Matter
  Phys.}\ }\textbf {\bibinfo {volume} {2}},\ \bibinfo {pages} {329} (\bibinfo
  {year} {2011})}\BibitemShut {NoStop}%
\bibitem [{\citenamefont {den Broeck}\ and\ \citenamefont
  {Esposito}(2015)}]{VanDenBroeck2015}%
  \BibitemOpen
  \bibfield  {author} {\bibinfo {author} {\bibfnamefont {C.~V.}\ \bibnamefont
  {den Broeck}}\ and\ \bibinfo {author} {\bibfnamefont {M.}~\bibnamefont
  {Esposito}},\ }\href {\doibase https://doi.org/10.1016/j.physa.2014.04.035}
  {\bibfield  {journal} {\bibinfo  {journal} {Physica A}\ }\textbf {\bibinfo
  {volume} {418}},\ \bibinfo {pages} {6 } (\bibinfo {year} {2015})}\BibitemShut
  {NoStop}%
\bibitem [{\citenamefont {Parrondo}\ \emph {et~al.}(2015)\citenamefont
  {Parrondo}, \citenamefont {Horowitz},\ and\ \citenamefont
  {Sagawa}}]{Parrondo2015}%
  \BibitemOpen
  \bibfield  {author} {\bibinfo {author} {\bibfnamefont {J.~M.~R.}\
  \bibnamefont {Parrondo}}, \bibinfo {author} {\bibfnamefont {J.~M.}\
  \bibnamefont {Horowitz}}, \ and\ \bibinfo {author} {\bibfnamefont
  {T.}~\bibnamefont {Sagawa}},\ }\href {https://doi.org/10.1038/nphys3230}
  {\bibfield  {journal} {\bibinfo  {journal} {Nat. Phys.}\ }\textbf {\bibinfo
  {volume} {11}},\ \bibinfo {pages} {131} (\bibinfo {year} {2015})}\BibitemShut
  {NoStop}%
\bibitem [{\citenamefont {de~Groot}\ and\ \citenamefont
  {Mazur}(1984)}]{Groot1984}%
  \BibitemOpen
  \bibfield  {author} {\bibinfo {author} {\bibfnamefont {S.~R.}\ \bibnamefont
  {de~Groot}}\ and\ \bibinfo {author} {\bibfnamefont {P.}~\bibnamefont
  {Mazur}},\ }\href@noop {} {\emph {\bibinfo {title} {Non-Equilibrium
  Thermodynamics}}}\ (\bibinfo  {publisher} {Dover},\ \bibinfo {year}
  {1984})\BibitemShut {NoStop}%
\bibitem [{\citenamefont {Glansdorff}\ and\ \citenamefont
  {Prigogine}(1971)}]{Glansdorff1971}%
  \BibitemOpen
  \bibfield  {author} {\bibinfo {author} {\bibfnamefont {P.}~\bibnamefont
  {Glansdorff}}\ and\ \bibinfo {author} {\bibfnamefont {I.}~\bibnamefont
  {Prigogine}},\ }\href@noop {} {\emph {\bibinfo {title} {Thermodynamic theory
  of structure, stability and fluctuations}}}\ (\bibinfo  {publisher}
  {Wiley-Interscience},\ \bibinfo {year} {1971})\BibitemShut {NoStop}%
\bibitem [{\citenamefont {Polettini}\ and\ \citenamefont
  {Esposito}(2014)}]{Polettini2014}%
  \BibitemOpen
  \bibfield  {author} {\bibinfo {author} {\bibfnamefont {M.}~\bibnamefont
  {Polettini}}\ and\ \bibinfo {author} {\bibfnamefont {M.}~\bibnamefont
  {Esposito}},\ }\href {\doibase 10.1063/1.4886396} {\bibfield  {journal}
  {\bibinfo  {journal} {J. Chem. Phys.}\ }\textbf {\bibinfo {volume} {141}},\
  \bibinfo {pages} {024117} (\bibinfo {year} {2014})}\BibitemShut {NoStop}%
\bibitem [{\citenamefont {Rao}\ and\ \citenamefont
  {Esposito}(2018{\natexlab{b}})}]{Rao2018a}%
  \BibitemOpen
  \bibfield  {author} {\bibinfo {author} {\bibfnamefont {R.}~\bibnamefont
  {Rao}}\ and\ \bibinfo {author} {\bibfnamefont {M.}~\bibnamefont {Esposito}},\
  }\href {\doibase 10.1088/1367-2630/aaa15f} {\bibfield  {journal} {\bibinfo
  {journal} {New J. Phys.}\ }\textbf {\bibinfo {volume} {20}},\ \bibinfo
  {pages} {023007} (\bibinfo {year} {2018}{\natexlab{b}})}\BibitemShut
  {NoStop}%
\bibitem [{\citenamefont {Fisher}(1937)}]{Fisher1937}%
  \BibitemOpen
  \bibfield  {author} {\bibinfo {author} {\bibfnamefont {R.~A.}\ \bibnamefont
  {Fisher}},\ }\href {\doibase 10.1111/j.1469-1809.1937.tb02153.x} {\bibfield
  {journal} {\bibinfo  {journal} {Ann. Eugenics}\ }\textbf {\bibinfo {volume}
  {7}},\ \bibinfo {pages} {355} (\bibinfo {year} {1937})}\BibitemShut {NoStop}%
\bibitem [{\citenamefont {Prigogine}\ and\ \citenamefont
  {Lefever}(1968)}]{Prigogine1968}%
  \BibitemOpen
  \bibfield  {author} {\bibinfo {author} {\bibfnamefont {I.}~\bibnamefont
  {Prigogine}}\ and\ \bibinfo {author} {\bibfnamefont {R.}~\bibnamefont
  {Lefever}},\ }\href {\doibase 10.1063/1.1668896} {\bibfield  {journal}
  {\bibinfo  {journal} {J. Chem. Phys.}\ }\textbf {\bibinfo {volume} {48}},\
  \bibinfo {pages} {1695} (\bibinfo {year} {1968})}\BibitemShut {NoStop}%
\bibitem [{\citenamefont {Auchmuty}\ and\ \citenamefont
  {Nicolis}(1976)}]{Auchmuty1976}%
  \BibitemOpen
  \bibfield  {author} {\bibinfo {author} {\bibfnamefont {J.~F.~G.}\
  \bibnamefont {Auchmuty}}\ and\ \bibinfo {author} {\bibfnamefont
  {G.}~\bibnamefont {Nicolis}},\ }\href
  {https://doi.org/10.1016/S0092-8240(77)90012-X} {\bibfield  {journal}
  {\bibinfo  {journal} {Bull. Math. Biol.}\ }\textbf {\bibinfo {volume} {38}},\
  \bibinfo {pages} {325} (\bibinfo {year} {1976})}\BibitemShut {NoStop}%
\bibitem [{\citenamefont {Peka\v{r}}(2005)}]{Pekar2005}%
  \BibitemOpen
  \bibfield  {author} {\bibinfo {author} {\bibfnamefont {M.}~\bibnamefont
  {Peka\v{r}}},\ }\href {\doibase doi:10.3184/007967405777874868} {\bibfield
  {journal} {\bibinfo  {journal} {Prog. React. Kinet. Mech.}\ }\textbf
  {\bibinfo {volume} {30}},\ \bibinfo {pages} {3} (\bibinfo {year}
  {2005})}\BibitemShut {NoStop}%
\bibitem [{\citenamefont {Ge}\ and\ \citenamefont {Qian}(2016)}]{Ge2016}%
  \BibitemOpen
  \bibfield  {author} {\bibinfo {author} {\bibfnamefont {H.}~\bibnamefont
  {Ge}}\ and\ \bibinfo {author} {\bibfnamefont {H.}~\bibnamefont {Qian}},\
  }\href {\doibase https://doi.org/10.1016/j.chemphys.2016.03.026} {\bibfield
  {journal} {\bibinfo  {journal} {Chem. Phys.}\ }\textbf {\bibinfo {volume}
  {472}},\ \bibinfo {pages} {241 } (\bibinfo {year} {2016})}\BibitemShut
  {NoStop}%
\bibitem [{\citenamefont {Fermi}(1956)}]{Fermi1956}%
  \BibitemOpen
  \bibfield  {author} {\bibinfo {author} {\bibfnamefont {E.}~\bibnamefont
  {Fermi}},\ }\href@noop {} {\emph {\bibinfo {title} {Thermodynamics}}}\
  (\bibinfo  {publisher} {Dover},\ \bibinfo {address} {New York},\ \bibinfo
  {year} {1956})\BibitemShut {NoStop}%
\bibitem [{\citenamefont {Cover}\ and\ \citenamefont
  {Thomas}(2012)}]{cover2012}%
  \BibitemOpen
  \bibfield  {author} {\bibinfo {author} {\bibfnamefont {T.~M.}\ \bibnamefont
  {Cover}}\ and\ \bibinfo {author} {\bibfnamefont {J.~A.}\ \bibnamefont
  {Thomas}},\ }\href@noop {} {\emph {\bibinfo {title} {Elements of Information
  Theory}}}\ (\bibinfo  {publisher} {Wiley},\ \bibinfo {year}
  {2012})\BibitemShut {NoStop}%
\bibitem [{\citenamefont {Auchmuty}\ and\ \citenamefont
  {Nicolis}(1975)}]{Auchmuty1975}%
  \BibitemOpen
  \bibfield  {author} {\bibinfo {author} {\bibfnamefont {J.}~\bibnamefont
  {Auchmuty}}\ and\ \bibinfo {author} {\bibfnamefont {G.}~\bibnamefont
  {Nicolis}},\ }\href {\doibase https://doi.org/10.1016/S0092-8240(75)80036-X}
  {\bibfield  {journal} {\bibinfo  {journal} {Bull. Math. Biol.}\ }\textbf
  {\bibinfo {volume} {37}},\ \bibinfo {pages} {323} (\bibinfo {year}
  {1975})}\BibitemShut {NoStop}%
\bibitem [{\citenamefont
  {Herschkowitz-Kaufman}(1975)}]{Herschkowitz-Kaufman1975}%
  \BibitemOpen
  \bibfield  {author} {\bibinfo {author} {\bibfnamefont {M.}~\bibnamefont
  {Herschkowitz-Kaufman}},\ }\href {\doibase 10.1007/BF02459527} {\bibfield
  {journal} {\bibinfo  {journal} {Bull. Math. Biol.}\ }\textbf {\bibinfo
  {volume} {37}},\ \bibinfo {pages} {589} (\bibinfo {year} {1975})}\BibitemShut
  {NoStop}%
\bibitem [{\citenamefont {Kolchinsky}\ and\ \citenamefont
  {Wolpert}(2018)}]{Kolchinsky2018}%
  \BibitemOpen
  \bibfield  {author} {\bibinfo {author} {\bibfnamefont {A.}~\bibnamefont
  {Kolchinsky}}\ and\ \bibinfo {author} {\bibfnamefont {D.~H.}\ \bibnamefont
  {Wolpert}},\ }\href {\doibase 10.1098/rsfs.2018.0041} {\bibfield  {journal}
  {\bibinfo  {journal} {Interface Focus}\ }\textbf {\bibinfo {volume} {8}},\
  \bibinfo {pages} {20180041} (\bibinfo {year} {2018})}\BibitemShut {NoStop}%
\bibitem [{\citenamefont {Rossi}\ \emph {et~al.}(2008)\citenamefont {Rossi},
  \citenamefont {Ristori}, \citenamefont {Rustici}, \citenamefont
  {Marchettini},\ and\ \citenamefont {Tiezzi}}]{Rossi2008}%
  \BibitemOpen
  \bibfield  {author} {\bibinfo {author} {\bibfnamefont {F.}~\bibnamefont
  {Rossi}}, \bibinfo {author} {\bibfnamefont {S.}~\bibnamefont {Ristori}},
  \bibinfo {author} {\bibfnamefont {M.}~\bibnamefont {Rustici}}, \bibinfo
  {author} {\bibfnamefont {N.}~\bibnamefont {Marchettini}}, \ and\ \bibinfo
  {author} {\bibfnamefont {E.}~\bibnamefont {Tiezzi}},\ }\href {\doibase
  https://doi.org/10.1016/j.jtbi.2008.08.026} {\bibfield  {journal} {\bibinfo
  {journal} {J. Theor. Biol.}\ }\textbf {\bibinfo {volume} {255}},\ \bibinfo
  {pages} {404 } (\bibinfo {year} {2008})}\BibitemShut {NoStop}%
\bibitem [{\citenamefont {Padirac}\ \emph {et~al.}(2013)\citenamefont
  {Padirac}, \citenamefont {Fujii}, \citenamefont {Est{\'e}vez-Torres},\ and\
  \citenamefont {Rondelez}}]{Padirac2013}%
  \BibitemOpen
  \bibfield  {author} {\bibinfo {author} {\bibfnamefont {A.}~\bibnamefont
  {Padirac}}, \bibinfo {author} {\bibfnamefont {T.}~\bibnamefont {Fujii}},
  \bibinfo {author} {\bibfnamefont {A.}~\bibnamefont {Est{\'e}vez-Torres}}, \
  and\ \bibinfo {author} {\bibfnamefont {Y.}~\bibnamefont {Rondelez}},\ }\href
  {\doibase 10.1021/ja403584p} {\bibfield  {journal} {\bibinfo  {journal} {J.
  Am. Chem. Soc.}\ }\textbf {\bibinfo {volume} {135}},\ \bibinfo {pages}
  {14586} (\bibinfo {year} {2013})}\BibitemShut {NoStop}%
\bibitem [{\citenamefont {Tayar}\ \emph {et~al.}(2015)\citenamefont {Tayar},
  \citenamefont {Karzbrun}, \citenamefont {Noireaux},\ and\ \citenamefont
  {Bar-Ziv}}]{Tayar2015}%
  \BibitemOpen
  \bibfield  {author} {\bibinfo {author} {\bibfnamefont {A.~M.}\ \bibnamefont
  {Tayar}}, \bibinfo {author} {\bibfnamefont {E.}~\bibnamefont {Karzbrun}},
  \bibinfo {author} {\bibfnamefont {V.}~\bibnamefont {Noireaux}}, \ and\
  \bibinfo {author} {\bibfnamefont {R.~H.}\ \bibnamefont {Bar-Ziv}},\ }\href
  {https://doi.org/10.1038/nphys3469} {\bibfield  {journal} {\bibinfo
  {journal} {Nat. Phys.}\ }\textbf {\bibinfo {volume} {11}},\ \bibinfo {pages}
  {1037} (\bibinfo {year} {2015})}\BibitemShut {NoStop}%
\bibitem [{\citenamefont {Adamatzky}\ \emph {et~al.}(2005)\citenamefont
  {Adamatzky}, \citenamefont {De~Lacy~Costello},\ and\ \citenamefont
  {Asai}}]{Adamatzky2005}%
  \BibitemOpen
  \bibfield  {author} {\bibinfo {author} {\bibfnamefont {A.}~\bibnamefont
  {Adamatzky}}, \bibinfo {author} {\bibfnamefont {B.}~\bibnamefont
  {De~Lacy~Costello}}, \ and\ \bibinfo {author} {\bibfnamefont
  {T.}~\bibnamefont {Asai}},\ }\href@noop {} {\emph {\bibinfo {title}
  {Reaction-diffusion computers}}}\ (\bibinfo  {publisher} {Elsevier},\
  \bibinfo {year} {2005})\BibitemShut {NoStop}%
\bibitem [{\citenamefont {Sartori}\ \emph {et~al.}(2014)\citenamefont
  {Sartori}, \citenamefont {Granger}, \citenamefont {Lee},\ and\ \citenamefont
  {Horowitz}}]{Sartori2014}%
  \BibitemOpen
  \bibfield  {author} {\bibinfo {author} {\bibfnamefont {P.}~\bibnamefont
  {Sartori}}, \bibinfo {author} {\bibfnamefont {L.}~\bibnamefont {Granger}},
  \bibinfo {author} {\bibfnamefont {C.~F.}\ \bibnamefont {Lee}}, \ and\
  \bibinfo {author} {\bibfnamefont {J.~M.}\ \bibnamefont {Horowitz}},\ }\href
  {\doibase 10.1371/journal.pcbi.1003974} {\bibfield  {journal} {\bibinfo
  {journal} {PLOS Comp. Biol.}\ }\textbf {\bibinfo {volume} {10}},\ \bibinfo
  {pages} {1} (\bibinfo {year} {2014})}\BibitemShut {NoStop}%
\bibitem [{\citenamefont {Barato}\ \emph {et~al.}(2014)\citenamefont {Barato},
  \citenamefont {Hartich},\ and\ \citenamefont {Seifert}}]{Barato2014}%
  \BibitemOpen
  \bibfield  {author} {\bibinfo {author} {\bibfnamefont {A.~C.}\ \bibnamefont
  {Barato}}, \bibinfo {author} {\bibfnamefont {D.}~\bibnamefont {Hartich}}, \
  and\ \bibinfo {author} {\bibfnamefont {U.}~\bibnamefont {Seifert}},\ }\href
  {\doibase 10.1088/1367-2630/16/10/103024} {\bibfield  {journal} {\bibinfo
  {journal} {New J. Phys.}\ }\textbf {\bibinfo {volume} {16}},\ \bibinfo
  {pages} {103024} (\bibinfo {year} {2014})}\BibitemShut {NoStop}%
\bibitem [{\citenamefont {Ito}\ and\ \citenamefont {Sagawa}(2015)}]{Ito2015}%
  \BibitemOpen
  \bibfield  {author} {\bibinfo {author} {\bibfnamefont {S.}~\bibnamefont
  {Ito}}\ and\ \bibinfo {author} {\bibfnamefont {T.}~\bibnamefont {Sagawa}},\
  }\href {https://doi.org/10.1038/ncomms8498} {\bibfield  {journal} {\bibinfo
  {journal} {Nat. Commun.}\ }\textbf {\bibinfo {volume} {6}},\ \bibinfo {pages}
  {7498} (\bibinfo {year} {2015})}\BibitemShut {NoStop}%
\bibitem [{\citenamefont {Falcke}(2003)}]{Falcke2003}%
  \BibitemOpen
  \bibfield  {author} {\bibinfo {author} {\bibfnamefont {M.}~\bibnamefont
  {Falcke}},\ }\href {\doibase 10.1088/1367-2630/5/1/396} {\bibfield  {journal}
  {\bibinfo  {journal} {New J. Phys.}\ }\textbf {\bibinfo {volume} {5}},\
  \bibinfo {pages} {96} (\bibinfo {year} {2003})}\BibitemShut {NoStop}%
\bibitem [{\citenamefont {Skupin}\ \emph {et~al.}(2010)\citenamefont {Skupin},
  \citenamefont {Kettenmann},\ and\ \citenamefont {Falcke}}]{Skupin2010}%
  \BibitemOpen
  \bibfield  {author} {\bibinfo {author} {\bibfnamefont {A.}~\bibnamefont
  {Skupin}}, \bibinfo {author} {\bibfnamefont {H.}~\bibnamefont {Kettenmann}},
  \ and\ \bibinfo {author} {\bibfnamefont {M.}~\bibnamefont {Falcke}},\ }\href
  {\doibase 10.1371/journal.pcbi.1000870} {\bibfield  {journal} {\bibinfo
  {journal} {PLOS Comp. Biol.}\ }\textbf {\bibinfo {volume} {6}},\ \bibinfo
  {pages} {1} (\bibinfo {year} {2010})}\BibitemShut {NoStop}%
\end{thebibliography}%

\end{document}